# Electronic Structure of the Cuprate Superconducting and Pseudogap Phases from Spectroscopic Imaging STM

A R Schmidt [1,2,3 §], K Fujita [1,2,5], E -A Kim [1], M J Lawler [1,6], H Eisaki [7], S Uchida [5], D-H Lee [4] and  J C Davis [1,2,8]

1. LASSP, Department of Physics, Cornell University, Ithaca, NY 14853, USA
2. CMPMS Department, Brookhaven National Laboratory, Upton, NY 11973, USA
3. Quantum Nanoelectronics Laboratory, Department of Physics, University of California, Berkeley, California 94720, USA.
4. Department of Physics, University of California, Berkeley, California 94720, USA.
5. Department of Physics, University of Tokyo, Bunkyo-ku, Tokyo 113-0033, Japan
6. Department of Physics & Astronomy, Binghamton University, Binghamton, NY 13902, USA
7. Institute of Advanced Industrial Science and Technology, Tsukuba, Ibaraki 305-8568, Japan.
8. School of Physics & Astronomy, U. of St. Andrews, St. Andrews, Fife KY16 9SS, Scotland
§ Corresponding author; e-mail: andy.schmidt@berkeley.edu

**Abstract**. We survey the use of spectroscopic imaging STM to probe the electronic structure of underdoped cuprates. Two distinct classes of electronic states are observed in both the d-wave superconducting (dSC) and the pseudogap (PG) phases. The first class consists of the dispersive Bogoliubov quasiparticle excitations of a homogeneous d-wave superconductor, existing below a lower energy scale $E=\Delta_0$. We find that the Bogoliubov quasiparticle interference signatures of delocalized Cooper pairing are restricted to a **k**-space arc which terminates near the lines connecting $\mathbf{k}=\pm(\pi/a_0,0)$ to $\mathbf{k}=\pm(0,\pi/a_0)$. This arc shrinks continuously with decreasing hole density such that Luttinger's theorem could be satisfied if it represents the front side of a hole-pocket which is bounded behind by the lines between $\mathbf{k}=\pm(\pi/a_0,0)$ and $\mathbf{k}=\pm(0,\pi/a_0)$.  In both phases the only broken symmetries detected for the $|E|<\Delta_0$ states are those of a d-wave superconductor. The second class of states occurs proximate to the pseudogap energy scale $E=\Delta_1$. Here the non-dispersive electronic structure breaks the expected 90°-rotational symmetry of electronic structure within each unit cell, at least down to 180°-rotational symmetry. This Q=0 electronic symmetry breaking was first detected as an electronic inequivalence at the two oxygen sites within each unit cell by using a measure of nematic ($C_2$) symmetry.  Incommensurate non-dispersive conductance modulations, locally breaking both rotational and translational symmetries, coexist with this intra-unit-cell electronic symmetry breaking at $E=\Delta_1$. Their characteristic wavevector **Q** is determined by the **k**-space points where Bogoliubov quasiparticle interference terminates and therefore changes continuously with doping. The distinct broken electronic symmetry states (Q=0 and finite Q) coexisting at $E \sim \Delta_1$ are found to be indistinguishable in the dSC and PG phases. We propose that the next challenge for SI-STM studies is to determine the relationship of the $E \sim \Delta_1$ broken symmetry electronic states to the pseudogap phase, and to the $E < \Delta_0$ states associated with Cooper pairing.

**CONTENTS**

**1.       Basic Electronic Structure of Hole-doped Cuprates**

*1.1        Electronic Structure of the Superconducting and Pseudogap Phases*

The electronic structure of the $CuO_2$ plane is dominated by Cu *3d* and O *2p* orbitals [1]. Energetically each Cu $d_{x2-y2}$ orbital is split into singly and doubly occupied configurations by on-site Coulomb interactions, with the O *p*-states intervening. This is a 'charge-transfer' [1] Mott insulator which is strongly antiferromagnetic due to superexchange [2,3]. 'Hole-doping' is achieved by removing electrons from the O atoms [4]. It results in the highest temperature superconductivity available today. The phase diagram [5], with *p* the number of holes per $CuO_2$, is shown schematically in Figure 1a. Antiferromagnetism persists for *p* < 2-5%, superconductivity occurs in the range 5-10% < *p* < 25-30%, and a metallic state exists for *p* > 25-30%. The highest superconducting critical temperature $T_c$ always occurs at 'optimal doping' near *p*~16% and the superconductivity always exhibits d-wave symmetry. With reduced *p*, an electronic excitation with energy scale $E=\Delta_1$ that is anisotropic in **k**-space [5-10] appears at *T\** far above the superconducting $T_c$. This region is labelled the 'pseudogap' (PG) phase because $\Delta_1$ could be the energy gap of another ordered phase. Explanations for the PG phase include (i) that it occurs because of effects of a spin-liquid created by hole-doping the antiferromagnetic Mott insulator [3,11-15] or, (ii) that it is a phase incoherent d-wave superconductor [16-21] or, (iii) that it is an electronic ordered phase [22-36] due to the breaking of electronic symmetries unrelated to superconductivity. Another logically valid possibility is also that some combination of these effects is at play. A key challenge for cuprate studies is therefore to achieve a widely accepted understanding of the electronic structure of the PG phase, and to determine its relationship to the high temperature superconductivity.

*1.2    Two Characteristic Types of Electronic States in Underdoped Cuprates*

In underdoped cuprates, a variety of different spectroscopies reveal the two energy scales $\Delta_1$ and $\Delta_0$ in association with two distinct types of electronic excited states [5-7, 37-40]. The energies $\Delta_0$ and $\Delta_1$ diverge from one another with diminishing $p$ as shown in Fig. 1b (reproduced from [7]). ARPES (angle resolved photoemission) reveals that, in the PG phase, excitations with $E \sim \Delta_1$ occur in the regions of momentum space near $\mathbf{k} \cong (\pi/a_0,0); (0,\pi/a_0)$ and that $\Delta_1(p)$ increases rapidly as $p \to 0$ [6-9]. In contrast, the 'nodal' region of $\mathbf{k}$-space exhibits an ungapped 'Fermi Arc' [41] in the PG phase, and a momentum- and temperature-dependent energy gap opens upon this arc in the dSC phase [41-47]. Results from many other spectroscopies appear to be in agreement with this picture. For example, optical transient grating spectroscopy finds that the excitations near $\Delta_1$ propagate very slowly without recombination to form Cooper pairs, whereas lower energy excitations near the d-wave nodes propagate easily and reform delocalized Cooper pairs as expected [37]. Andreev tunneling exhibits two distinct excitation energy scales which diverge as $p \to 0$: the first is identified with the pseudogap energy $\Delta_1$ and the second lower scale $\Delta_0$ with the maximum pairing gap energy of delocalized Cooper-pairs [38]. Raman spectroscopy finds that scattering near the node is consistent with delocalized Cooper pairing whereas scattering at the antinodes is not [39]. Finally, muon spin rotation studies of the superfluid density show its evolution to be inconsistent with states on the whole Fermi surface being available for condensation, as if anti-nodal regions cannot contribute to delocalized Cooper pairs. [40]

Tunneling density-of-states measurements have established an energetically particle-hole symmetric excitation energy $E=\pm\Delta_1$ which is indistinguishable in magnitude in the PG and dSC phases [48, 49]. In Fig. 2b we show the evolution of spatially-averaged differential tunnelling conductance [50, 51, 52] $g(E)$ for $Bi_2Sr_2CaCu_2O_{8+\delta}$. The $p$-dependence of this pseudogap energy $E=\pm\Delta_1$ is indicated with blue dashed curve (see Sections 3, 5 and 7) while the approximate $p$-dependence of $\Delta_0$ (as determined from several different tunnelling techniques discussed in Section 3) is shown by red dashed curves.

The rich spatial complexity of the two classes of electronic states in underdoped cuprates has been exposed more recently by spectroscopic imaging scanning tunneling microscopy (SI-STM). For energies below the weakly doping dependent [52, 58] lower scale $E \sim \Delta_0$, the characteristics of dispersive Bogoliubov quasiparticles of a spatially homogeneous superconductor (Fig. 2a) are observed [53-59]. By contrast, the states near $E \sim \Delta_1$ are spatially disordered on the nm scale [50, 51, 52, 60-67]. More importantly, when the spatial structure of these non-dispersive states surrounding $E \sim \Delta_1$ is imaged with sub-angstrom precision, several distinct broken spatial symmetries are observed [50, 57, 58, 59, 68] (Fig. 2c). These two classes of excitations also exhibit increasing energy segregation in SI-STM data as $p \to 0$.

## 2      Spectroscopic Imaging Scanning Tunneling Microscopy

*2.1      Techniques and Challenges of SI-STM*

Imaging the differential tunneling conductance $dI/dV(\mathbf{r}, E = eV) \equiv g(\mathbf{r}, E = eV)$ with atomic resolution and register, and as a function of both location $\mathbf{r}$ and electron energy $E$, is referred to as spectroscopic imaging STM. This technique is distinct from other electron spectroscopies in that it can access simultaneously the real space ($\mathbf{r}$-space) and momentum space ($\mathbf{k}$-space) electronic structure for both filled and empty states. However, great care must be taken to avoid the serious systematic errors that are endemic to it, especially in the study of underdoped cuprates.

The first systematic problem occurs because the STM tip-sample tunneling current is given by

$$I(\mathbf{r}, z, V) = f(\mathbf{r}, z) \int_0^{eV} N(\mathbf{r}, E) dE \qquad (1)$$

where $z$ is the tip-surface distance, $V$ the tip-sample bias voltage, $N(\mathbf{r}, E)$ the sample's local-density-of-electronic-states, while $f(\mathbf{r}, z)$ contains effects of tip elevation and of spatially-dependent tunneling matrix elements. The $g(\mathbf{r}, E)$ data are then related to $N(\mathbf{r}, E)$ by [55, 57-59, 60]

$$g(\mathbf{r}, E = eV) = \frac{eI_S}{\int_0^{eV_S} N(\mathbf{r}, E') dE'} N(\mathbf{r}, E) \qquad (2)$$

where $V_S$ and $I_S$ are the (constant) junction 'set-up' bias voltage and current respectively. From Eqn. 2 we see that when $\int_0^{eVs} N(\mathbf{r}, E')dE'$ is strongly heterogeneous at the atomic scale (as it is typically in underdoped cuprates [50, 51, 52, 56-68]) $g(\mathbf{r}, E = eV)$ cannot be used to measure $N(\mathbf{r}, E)$. However, these potentially severe systematic errors can be cancelled [55, 57-59] by using the observable [55]

$$Z(\mathbf{r}, E) \equiv \frac{g(\mathbf{r}, E = +eV)}{g(\mathbf{r}, E = -eV)} = \frac{N(\mathbf{r}, +E)}{N(\mathbf{r}, -E)} \qquad (3)$$

which measures correctly the ratio of the density-of-states for electron injection to that for extraction at a given **r** and E. A related observable which also avoids these systematic errors (but lacks energy resolution) is [57]

$$R(\mathbf{r}) \equiv \frac{I(\mathbf{r}, E = +eV)}{I(\mathbf{r}, E = -eV)} = \frac{\int_0^{+eV} N(\mathbf{r}, E)dE}{\int_{-eV}^{0} N(\mathbf{r}, E)dE} \qquad (4)$$

A different challenge is the random nanoscale variation of $\Delta_1(\mathbf{r})$ which causes the $E \sim \Delta_1$ pseudogap states to be detected at different locations for different bias voltages (Fig. 3a). This problem can be mitigated [58, 68] by scaling the tunnel-bias energy $E=eV$ at each **r** by the pseudogap magnitude $\Delta_1(\mathbf{r})$ at the same location. This procedure defines a reduced energy scale $e = E/\Delta_1(\mathbf{r})$ such that

$$Z(\mathbf{r}, e) \equiv Z(\mathbf{r}, E/\Delta_1(\mathbf{r})) \qquad (5)$$

in which the $E \sim \Delta_1$ pseudogap states all occur together at $e=1$ [58].

Another important systematic error has to do with $g(\mathbf{q}, E)$ and $Z(\mathbf{q}, E)$, the Fourier transforms of $g(\mathbf{r}, E)$ and $Z(\mathbf{r}, E)$ respectively. These are used to distinguish any non-dispersive ordering wavevector $\mathbf{Q}^*$ of an electronic ordered phase from the dispersive wavevectors $\mathbf{q}_i(E)$ due to quantum interference patterns of delocalized states. But to achieve sufficient precision in $|\mathbf{q}_i(E)|$ for such discrimination requires that $g(\mathbf{r}, E)$ or $Z(\mathbf{r}, E)$ be measured in large fields-of-view (FOV) while maintaining atomic resolution and registry, and that the energy resolution be at or below ~2 meV. When any smaller FOV or poorer energy resolution is used in $g(\mathbf{r}, E)$ studies, the erroneous impression of non-dispersive modulations is created unavoidably. For $Bi_2Sr_2CaCu_2O_{8+\delta}$, we have demonstrated both empirically and based upon the principles of Fourier transformation that, in both the dSC and PG phases, no deductions distinguishing between dispersive and non-dispersive excitations can be made using Fourier transformed $g(\mathbf{r},E)$ data from a FOV smaller than ~45nm-square [54, 59].

*2.2  Systematic SI-STM Studies of $Bi_2Sr_2CaCu_2O_{8+\delta}$*

We have applied these techniques during the sequence of studies summarized herein by measuring $g(\mathbf{r},E)$ in ~ 45 nm square fields of view in $Bi_2Sr_2CaCu_2O_{8+\delta}$ samples with $p \cong 0.19, 0.17, 0.14, 0.12, 0.10, 0.08, 0.06$ or with $T_c(K) = 86, 88, 74, 64, 45, 37, 20$ respectively. Several of these samples were studied in both the dSC and PG phases. Each sample is inserted into the cryogenic ultra high vacuum of the SI-STM system, cleaved to reveal an atomically clean BiO surface, and all $g(\mathbf{r},E)$ measurements were made between 1.9 K and 65K. Three cryogenic SI-STM's (optimized for different purposes) are used throughout these studies. The resulting data set, acquired over approximately a decade, consists of $>10^8$ atomically resolved and registered tunneling spectra.

## 3  Nanoscale Electronic Disorder in $Bi_2Sr_2CaCu_2O_{8+\delta}$

*3.1  Nanoscale Electronic Disorder of the $E\sim\Delta_1$ Pseudogap States*

Nanoscale electronic disorder is universal in images of $\Delta_1(\mathbf{r})$ measured on $Bi_2Sr_2CaCu_2O_{8+\delta}$ samples [50-54, 57-68]. The values of $|\Delta_1|$ range from above 130meV to below 10meV as the hole-density $p$ ranges from 0.06 to 0.22. Highly similar nanoscale electronic disorder is seen in $Bi_2Sr_2Cu_1O_{6+\delta}$ [56, 65] and in $Bi_2Sr_2Ca_2Cu_3O_{10+\delta}$ [69]. In Figure 3a we show a typical $Bi_2Sr_2CaCu_2O_{8+\delta}$ $\Delta_1(\mathbf{r})$ image - upon which the sites of the non-stoichiometric oxygen dopant ions are overlaid [51]. Figure 3b shows the typical $g(E)$ spectrum associated with each different value of $\pm\Delta_1$ [50]. It also shows how the electronic structure becomes homogeneous [50-56, 58, 59] below a lower energy scale $E=\pm\Delta_o$ as indicated by the arrows. $Bi_2Sr_2CuO_{6+\delta}$ and $Bi_2Sr_2Ca_2Cu_3O_{10+\delta}$ samples show similar effects [56, 65, 69]. The distributions of $|\Delta_1|$ measured in units of the spatially-averaged $\overline{\Delta}_1$ from six samples with varying hole-densities are shown in Figure 3c. As these normalized distributions are virtually independent of $p$, the microscopic trigger for the $\Delta_1$-disorder appears universal. Imaging $\Delta_1(\mathbf{r})$ in the PG phase reveals highly similar [59, 64, 65, 66] nanoscale electronic disorder. Explaining these $\Delta_1$-disorder phenomena has been a fascinating challenge.

*3.2  Imaging the Effects of Interstitial Oxygen Dopant Atoms*

An important element of the explanation is that electron-acceptor atoms must be introduced [70] to generate hole-doped superconductivity from the Mott insulating phase. This almost always creates random distributions of differently charged dopant ions near the $CuO_2$ planes [71]. The dopant ions in $Bi_2Sr_2CaCu_2O_{8+\delta}$ are -2e charged interstitials and can conceivably cause a variety of different local effects. For example, electrostatic screening of each ion could accumulate holes at those locations thereby reducing the energy-gap values nearby [72, 73]. Or the dopant ions could generate local crystalline stress/strain [74-78] thereby disordering hopping matrix elements and electron-electron interactions within the unit cell. In $Bi_2Sr_2CaCu_2O_{8+\delta}$ the locations of interstitial dopant ions can be identified because an atomic scale impurity state occurs at E=-0.96V nearby each ion [51] (Fig. 3a). Strong spatial correlations are observed between the distribution of these impurity states and $\Delta_1(\mathbf{r})$ maps. This implies that dopant ion disorder is responsible for much of the $\Delta_1(\mathbf{r})$ electronic disorder. The primary effect near each dopant ion is a shift of spectral weight from low to high energy with the $\Delta_1$ excitation energy increasing strongly. Moreover, simultaneous imaging of the dopant ion locations and $g(\mathbf{r},E<\Delta_0)$ reveals that the dispersive $g(\mathbf{r},E)$ modulations due to scattering of Bogoliubov quasiparticles are well correlated with dopant ion locations meaning that the dopant ions are an important source of such scattering [50-56, 58, 59] (Sections 6 and 7). This demonstration that it is the chemical doping process itself which both disorders $\Delta_1$ and causes strong quasiparticle scattering is of significance because similar nanoscale electronic disorder phenomena are then likely to be common (although with different intensities) in all non-stoichiometric cuprates.

*3.3    Microscopic Mechanism of $\Delta_1$ Disorder*

The microscopic mechanism of the $\Delta_1$-disorder is not yet fully understood. Hole-accumulation surrounding $O^{2-}$ dopant ions does not appear to be the correct explanation because (i) the modulations in integrated density of filled states are observed to be weak [51] and (ii) $\Delta_1$ is increased nearby the dopant ions [51] a situation diametrically opposite to the expected effect from hole-accumulation there. Atomic substitution at random on the Sr site is known to suppress superconductivity strongly [71] possibly due to geometrical distortions of the unit cell and associated changes in the hopping matrix elements. It has therefore been proposed that the interstitial dopant ions might act similarly, perhaps by displacing the Sr or apical oxygen atoms [71, 74, 75] and thereby distorting the unit cell geometry. Direct support for this point of view comes from the observation that quasi-periodic distortions of the crystal unit-cell geometry yield virtually identical perturbations in g(E) and $\Delta_1(\mathbf{r})$ but now are unrelated to the dopant ions [79]. Thus it seems that the $\Delta_1$-disorder is not caused primarily by carrier density modulations but by geometrical distortions to the unit cell dimensions with resulting strong local changes in the high energy electronic structure.

*3.4    'Kinks' in g(E) separating Homogeneous and Heterogeneous States*

So-called "kinks" have been reported ubiquitously in cuprate g(E) spectra [50, 51, 52, 55, 56, 58-67]. In general, they are weak perturbations to N(E) near optimal doping, becoming more clear as *p* is diminished [50, 52]. Figure 3b demonstrates how, in $\Delta_1$-sorted $g(E)$ spectra, the kinks are universal but become more obvious for $\Delta_1 > 50$ meV [50, 52]. Each kink can be identified and its energy is labelled $\Delta_0(\mathbf{r})$. By determining $\overline{\Delta}_0$ (the spatial average of $\Delta_0(\mathbf{r})$) as a function of $p$, we find that this energy $\overline{\Delta}_0$ always divides the electronic structure into two categories [52]. For $E < \overline{\Delta}_0$ the excitations are homogenous in $\mathbf{r}$-space and well defined Bogoliubov quasiparticle eigenstates in $\mathbf{k}$-space (Section 6). By contrast, the pseudogap excitations at $E \sim \Delta_1$ are heterogeneous in $\mathbf{r}$-space and ill-defined in $\mathbf{k}$-space (Section 7). Figure 3c provides a summary of the evolution of $\overline{\Delta}_0$ and $\overline{\Delta}_1$ with $p$.

*3.5 Summary*

The $\Delta_1$-disorder of $Bi_2Sr_2CaCu_2O_{8+\delta}$ is strongly influenced by the random distribution of dopant ions [51]. This occurs through an electronic process in which geometrical distortions of the crystal unit cell play a prominent role [76-79]. The disorder is strongly reflected in the electronic excitations near the pseudogap energy $E \sim \Delta_1$. The electronic excitations with $E < \Delta_0$, in contrast, are only influenced by the dopant ions via scattering; they are otherwise relatively homogeneous when studied using QPI or by direct imaging [50, 51, 52, 61]. As the equivalent $\Delta_1(\mathbf{r})$ disorder is observed in the PG phase, [59, 64, 65, 66, 68], an appealing idea has been that these $\Delta_1(\mathbf{r})$ arrangements (Fig. 3a) represent images the superconducting 'grains' of a granular superconductor. However, the superconducting energy gap $\Delta(\mathbf{k})$ when determined using Bogoliubov QPI is deduced to be rather spatially homogeneous [50, 53-56, 58, 59]. Moreover, the $E \sim \Delta_1$ pseudogap states exhibit a classic oxygen isotope effect which indicates a strong localized electron-lattice interaction [80]. Finally, atomic resolution imaging of the $E \sim \Delta_1$ states shows them to be non-dispersive and to break several spatial symmetries locally [57, 58, 59, 68] (Section 7). As none of these latter phenomena are the predicted characteristics of d-wave Bogoliubov quasiparticles within a superconducting grain, it appears implausible at present that $\Delta_1(\mathbf{r})$ represents merely the image of a d-wave granular superconductor.

**4      Bogoliubov Quasiparticle Interference Imaging**

*4.1     d-Wave Bogoliubov Quasiparticle Interference*

Bogoliubov quasiparticles are the excitations generated by breaking Cooper pairs. Bogoliubov quasiparticle interference (QPI) occurs when these quasiparticle de Broglie waves are scattered by impurities and the scattered waves undergo quantum interference. In a d-wave cuprate-like superconductor with a single hole-like band of uncorrelated electrons, the Bogoliubov quasiparticle dispersion $E(\mathbf{k})$ would have 'banana-shaped' constant energy contours. For a given energy $E$, the d-symmetry of the superconducting energy gap would then cause strong maxima to appear in the joint-density-of-states at the eight tips $\mathbf{k}_j(E)$; $j = 1, 2,..., 8$ of these 'bananas'. Elastic scattering between the $\mathbf{k}_j(E)$ then produces $\mathbf{r}$-space interference patterns in the local-density-of-states $N(\mathbf{r},E)$. The resulting $g(\mathbf{r},E)$ modulations detectable by SI-STM should exhibit 16 ±$\mathbf{q}$ pairs of dispersive wavevectors in $g(\mathbf{q},E)$ (Fig. 4a). The set of these wavevectors that is specifically characteristic of d-wave superconductivity consists of seven: $\mathbf{q}_i(E)$ $i$=1,....,7 with $\mathbf{q}_i(-E) = \mathbf{q}_i(+E)$. This is the so-called 'octet model' [81-83] within which, by using the point-group symmetry of the first $CuO_2$ Brillouin zone, the locus of the above-mentioned tips at $\mathbf{k}_B(E) = (k_x(E), k_y(E))$ is determined from:

$$\begin{aligned}
\mathbf{q}_1 &= (2k_x, 0) & \mathbf{q}_4 &= (2k_x, 2k_y) & \mathbf{q}_7 &= (k_x - k_y, k_y - k_x) \\
\mathbf{q}_2 &= (k_x + k_y, k_y - k_x) & \mathbf{q}_5 &= (0, 2k_y) & & \\
\mathbf{q}_3 &= (k_x + k_y, k_y + k_x) & \mathbf{q}_6 &= (k_x - k_y, k_y + k_x) & &
\end{aligned} \qquad (6)$$

When these $\mathbf{q}_i(E)$ are measured from $Z(\mathbf{q},E)$, the Fourier transform of spatial modulations seen in $g(\mathbf{r},E)$ (see Fig. 2a for example), the $\mathbf{k}_B(E)$ can then be determined by using Eqn. 6 within the requirement that all its independent solutions be consistent at all energies. The superconductor's Cooper-pairing energy gap $\Delta(\mathbf{k})$ is then determined directly by inverting $\mathbf{k}_B(E=\Delta)$. In $Bi_2Sr_2CaCu_2O_{8+\delta}$ near optimal doping, measurements from QPI of the Fermi surface location $\mathbf{k}_B(E)$, and of the superconducting $\Delta(\mathbf{k})$ (Fig. 4b), are consistent with ARPES [54, 84]. In both $Ca_{2-x}Na_xCuO_2Cl_2$ and $Bi_2Sr_2CaCu_2O_{8+\delta}$ the QPI octet model yields $\mathbf{k}_B(E)$ and $\Delta(\mathbf{k})$ equally well [55, 56]. Moreover, the basic validity of the fundamental $\mathbf{k}$-space phenomenology behind the d-wave QPI 'octet' model has been confirmed by ARPES studies [85-87].

*4.2 Summary*

Fourier transformation of $Z(\mathbf{r},E)$ in combination with the octet model of d-wave Bogoliubov QPI yields the two branches of the Bogoliubov excitation spectrum $\mathbf{k}_B(\pm E)$ plus the superconducting energy gap magnitude $\pm\Delta(\mathbf{k})$ along the specific $\mathbf{k}$-space trajectory $\mathbf{k}_B$ for both filled and empty states in a single experiment. As only the Bogoliubov states of a d-wave superconductor could exhibit such a set of 16 pairs of interference wavevectors with $\mathbf{q}_i(-E)=\mathbf{q}_i(+E)$ and all dispersions internally consistent within the octet model, the energy gap $\pm\Delta(\mathbf{k})$ determined by these procedures is definitely that of the delocalized Cooper-pairs.

**5      Low Energy Excitations of the Superconducting Phase**

*5.1    Bogoliubov Quasiparticle Interference in the dSC Phase*

Bogoliubov QPI imaging techniques have been used to study the evolution of **k**-space electronic structure with falling $p$ in $Bi_2Sr_2CaCu_2O_{8+\delta}$. In the SC phase, the expected 16 pairs of **q**-vectors are always observed in $Z(\mathbf{q},E)$ and are found consistent with each other within the octet model (Fig. 2a, 4c). Remarkably, however, we find that in underdoped $Bi_2Sr_2CaCu_2O_{8+\delta}$ the dispersion of octet model **q**-vectors always stops at the same weakly doping-dependent [50, 56, 58] excitation energy $\Delta_0$ and at **q**-vectors indicating that the relevant **k**-space states are still far from the boundary of the Billouin zone. These observations are quite unexpected in the context of the d-wave BCS octet model. Moreover, for $E>\Delta_0$ the dispersive octet of **q**-vectors disappears and we observe three non-dispersive **q**-vectors: the reciprocal lattice vector **Q** along with $\mathbf{q}_1^*$ and $\mathbf{q}_5^*$ (see Fig. 4c). The equivalent pair of non-dispersive wavevectors to $\mathbf{q}_1^*$ and $\mathbf{q}_5^*$ has also been detected by SI-STM in $Ca_{2-x}Na_xCuO_2Cl_2$ [42] and $Bi_2Sr_2Cu_1O_{6+\delta}$ [56], and by ARPES in $Ca_{2-x}Na_xCuO_2Cl_2$ [42] and $Bi_2Sr_2CaCu_2O_{8+\delta}$ [86,87].

By using the QPI imaging techniques described in Section 4, we show in Fig. 4d the locus of Bogoliubov quasiparticle states $\mathbf{k}_B(E)$ determined as a function of $p$. Here we see that when the Bogoliubov QPI patterns disappear at $\Delta_0$, the **k**-states are near the diagonal lines between $\mathbf{k}=(0, \pi/a_0)$ and $\mathbf{k}=(\pi/a_0,0)$ within the $CuO_2$ Brillouin zone. These **k**-space Bogoliubov arc tips are defined by both the change from dispersive to non-dispersive characteristics and by the disappearance of the $\mathbf{q}_2$, $\mathbf{q}_3$, $\mathbf{q}_6$ and $\mathbf{q}_7$ modulations (see Fig. 4c). Thus, the signature of delocalized Cooper pairing is confined to an arc (fine solid lines in Fig. 4d) and this arc shrinks with falling $p$ [58]. This discovery has been supported directly by angle resolved photoemission studies [40,47] and by SI-STM studies of $Ca_{2-x}Na_xCuO_2Cl_2$ [55] and $Bi_2Sr_2Cu_1O_{6+\delta}$ [56], and indirectly by analyses of $g(\mathbf{r},E)$ by fitting to a multi-parameter model for **k**-space structure in the presence of a dSC energy gap [67].

The minima (maxima) of the Bogoliubov bands $\mathbf{k}_B(\pm E)$ should occur at the **k**-space location of the Fermi surface of the non-superconducting state. One can therefore ask if the carrier-density count satisfies Luttinger's theorem, which states that twice the **k**-space area enclosed by the Fermi surface, measured in units of the area of the first Brillouin zone, equals the number of electrons per unit cell, $n$. In Fig. 4d we show as fine solid lines hole-like Fermi surfaces fitted to our measured $\mathbf{k}_B(E)$. Using Luttinger's theorem with these **k**-space contours extended to the zone face would result in a calculated hole-density $p$ for comparison with the estimated hole density in the samples. These data are shown by filled symbols in the inset to Fig. 4d. We see that the Luttinger theorem is strongly violated at all doping below $p\sim10\%$. However, the Luttinger theorem can be amended in a doped Mott insulator [58] so that the zero-energy contours bounding the region representing carriers are defined, not only by poles in the Green's functions, but also by their zeros [*88*]. The locus of zeros of these Green's functions could be expected to occur at the lines joining **k**=(0, $\pi/a_0$) to **k**=($\pi/a_0$,0). In that situation, the hole density is related quantitatively to the area between the **k**=(0, $\pi/a_0$) - **k**=($\pi/a_0$,0) lines and the arcs. The carrier densities calculated in this fashion are shown by open symbols in the inset to Fig. 4d and are obviously in much better agreement with the chemical hole-density.

Figure 5 provides a doping-dependence analysis of the locations of the ends of the arc-tips at which Bogoliubov QPI signature disappears and where the **q**$_1$* and **q**$_5$* non-dispersive modulations appear. Figure 5a shows a typical Z(**q**,E) for which $\Delta_0<E<\Delta_1$. Here the vectors **q**$_1$* and **q**$_5$* (see Fig. 4c) are labeled along with the Bragg vectors **Q**$_x$ and **Q**$_y$. Figure 5b shows a schematic representation of the arc of the **k**-space supporting Bogoliubov QPI in blue. We show below how its termination points on the lines linking **k** =±(0,$\pi/a_0$) and **k** =±($\pi/a_0$,0) directly link the **q**$_1$* and **q**$_5$* wavevectors to the CuO$_2$ Brillouin zone size (via the arrows shown in red). Fig. 5c shows the doping dependence for Bi$_2$Sr$_2$CaCu$_2$O$_{8+\delta}$ of the location of both **q**$_1$* and **q**$_5$* measured from Z(**q**,E) [58]. The measured magnitude of **q**$_1$* and **q**$_5$* versus *p* are then shown in Fig. 5d along with the sum **q**$_1$*+**q**$_5$* which is always equal to $2\pi$. This demonstrates that, as the Bogoliubov QPI extinction point travels along the line from **k**=(0,$\pi/a_0$) and **k**=($\pi/a_0$,0) ([58] and Fig. 4d, Fig. 5b), the wavelengths of incommensurate modulations **q**$_1$* and **q**$_5$* are controlled by its **k**-space location [58]. Equivalent phenomena have also been reported for Bi$_2$Sr$_2$CuO$_{6+\delta}$ [56].

*5.2    Summary*

Because the superconducting $\Delta(\mathbf{k})$ must be translationally invariant for Bogoliubov QPI to exhibit the observed ~long range interference patterns, cuprate superconductivity is found to be rather spatially homogeneous (as implied also by direct g(E) spectra studies [50,67]). When *p* is reduced, the Bogoliubov QPI signature of which **k**-space states contribute to Cooper pairing is confined to an arc [50,54,56,58] in **k**-space which shrinks with falling doping. The arc tips lie near the diagonal lines connecting **k**=(0,±$\pi/a_0$) and **k**=(±$\pi/a_0$,0) and occur at a weakly doping-dependent [50, 58] energy $E = \Delta_0$ that is indistinguishable from (i) where the g(E) kinks occur [50] and (ii) where electronic homogeneity is lost [50, 56, 58-61]. The shrinking of this arc with decreasing hole-density could satisfy Luttinger's theorem if it is actually the front side of a hole-pocket bounded behind by the **k**=(0,±$\pi/a_0$) - **k**=(±$\pi/a_0$,0) lines. We find that the gap energy at the arc tip $\Delta_0$ is associated with the disappearance of the QPI signature of delocalized Cooper pairs for $E \geq \Delta_0$ (and simultaneously also the loss of electronic homogeneity and the kink in the density of states), while the upper energy $\Delta_1$ is associated with a quite distinct **r**-space electronic structure of the $E\sim\Delta_1$ pseudogap excitations (Section 7). Finally, the wavelengths of incommensurate modulations **q**$_1$* and **q**$_5$* are controlled by the **k**-space locations at which the Bogoliubov QPI signatures disappear, and these points evolve continuously with doping along the line joining **k**=$(0,\pm\pi/a_0)$ - **k**=$(\pm\pi/a_0,0)$.

## 6      Low Energy Excitations in the Pseudogap Phase

*6.1      QPI in a Phase-Fluctuating d-Wave Superconductor*

Because cuprate superconductivity is quasi-two-dimensional, the superfluid density increases from zero approximately linearly with $p$, and the superconducting energy gap $\Delta(\mathbf{k})$ exhibits four **k**-space nodes, fluctuations of the quantum phase $\phi(\mathbf{r},t)$ of the superconducting order parameter $\Psi=\Delta(\mathbf{k})e^{i\phi(\mathbf{r},t)}$ could have strong effects on the superconductivity at low hole-density [16-21]. Phenomena indicative of phase fluctuating superconductivity are detectable for cuprates in particular regions of the phase diagram [89-94] as indicated by the region $T_c<T<T_*$ (Fig. 1a). The techniques involved include terahertz transport studies [89], the Nernst effect [90, 91], torque-magnetometry measurements [92], field dependence of the diamagnetism [93], and zero-bias conductance enhancement [94].

A spectroscopic signature of phase incoherent d-wave superconductivity in the PG phase could be the continued existence of the Bogoliubov-like QPI octet described in the previous two sections. This is because, if the quantum phase $\phi(\mathbf{r},t)$ is fluctuating while the energy gap magnitude $\Delta(\mathbf{k})$ remains largely unchanged, the particle-hole symmetric octet of high joint-density-of-states regions generating the QPI should continue to exist [95-97]. However, any gapped **k**-space regions supporting Bogoliubov-like QPI in the PG phase must then occur beyond the tips of the ungapped Fermi Arc [41].

*6.2      Bogoliubov-like Quasiparticle Interference in the PG Phase*

The temperature evolution of the Bogoliubov octet in $Z(\mathbf{q},E)$ was studied as a function of increasing temperature from the dSC phase into the PG phase using a 48nm square FOV and with sub-unit-cell resolution. Representative $Z(\mathbf{q},E)$ for six temperatures are shown in Fig. 6; the $\boldsymbol{q}_i(E)$ ($i$=1,2,…,7) characteristic of the superconducting octet model are observed to remain unchanged upon passing above $T_c$ to at least $T \sim 1.5T_c$. This demonstrates that the Bogoliubov-like QPI octet phenomenology exists in the cuprate PG phase (although it is generated by different regions of **k**-space, and thus different $\Delta(\mathbf{k})$, than in the same sample in the SC phase). Thus for the low-energy (E<35mV) excitations in the underdoped PG phase, the $\boldsymbol{q}_i(E)$ ($i$=1,2,…,7) characteristic of the octet model are preserved unchanged upon passing above $T_c$. Importantly, all seven $q_i(E)$ ($i$=1,2,…,7) modulation wavevectors which are dispersive in the dSC phase remain dispersive into the PG phase still consistent with the octet model [59]. The octet wavevectors also retain their particle-hole symmetry $q_i(+E) = q_i(-E)$ in the PG phase and the g(**r**,E) modulations occur in the same energy range and emanate from the same contour in **k**-space as those observed at lowest temperatures [59]. However, with increasing $T$ the particle-hole symmetric energy gap $\Delta(\mathbf{k})$ closes near the nodes, leaving behind a growing Fermi arc of gapless excitations (Section 8.1).

*6.3    Summary*

All the Bogoliubov QPI signatures detectable in the dSC phase survive virtually unchanged into the underdoped PG phase - up to at least $T\sim1.5T_c$ for strongly underdoped $Bi_2Sr_2CaCu_2O_{8+\delta}$ samples. Moreover, for $E<\Delta_0$ all seven dispersive $q_i(E)$ modulations characteristic of the octet model in the dSC phase remain dispersive in the PG phase. These observations rule out the existence for all $E \leq \Delta_0$ of non-dispersive $g(E)$ modulations at finite ordering wavevector **Q*** which would be indicative of a static electronic order (breaking translational symmetry). This conclusion is in agreement with the results of ARPES studies [85, 86]. Instead, the observed excitations are indistinguishable from the dispersive **k**-space eigenstates of a phase incoherent d-wave superconductor [59]. Thus the SI-STM picture of electronic structure in the strongly underdoped PG phase actually contains three elements: (i) the ungapped Fermi arc [41], (ii) the particle-hole symmetric gap $\Delta(\mathbf{k})$ of a phase incoherent superconductor [59], and (iii) the non-dispersive and locally symmetry breaking excitations at the $E\sim\Delta_1$ energy scale [50, 57, 58, 59, 68] (which remain completely unaltered upon the transition between the dSC and the PG phases [59, 68]). This three-component description of the electronic structure of the cuprate pseudogap phase (Fig. 10d) has recently been confirmed in detail by ARPES studies [98].

**7    Broken Spatial Symmetries of $E\sim\Delta_1$ States in both the dSC and PG Phases**

*7.1    Atomic-scale Imaging of the $E\sim\Delta_1$ Pseudogap States*

In general for underdoped cuprates, the electronic excitations in the pseudogap energy range $E\sim\Delta_1$ are observed to be highly anomalous. They are associated with a strong antinodal pseudogap in **k**-space [8,9], they exhibit slow dynamics without recombination to form Cooper pairs [37], their Raman characteristics appear distinct from expectations for a d-wave superconductor [39], and they appear not to contribute to superfluid density [40].

As described in Sections 5 and 6, underdoped cuprates exhibit an octet of dispersive Bogoliubov QPI wavevectors $\mathbf{q}_i(E)$, but only upon a limited and doping-dependent arc in $\mathbf{k}$-space. But these effects always disappear above $E \cong \Delta_0$ to be replaced by a spectrum of non-dispersive states [50, 56, 57, 58, 59, 68] surrounding the pseudogap energy $E \sim \Delta_1$ (Fig. 4c). The $Z(\mathbf{q},E>\Delta_0)$ modulations exhibit the two non-dispersive $\mathbf{q}$-vectors, $\mathbf{q}_1^*$ and $\mathbf{q}_5^*$, which evolve with $p$ as shown in Fig. 5. The $\mathbf{q}_1^*$ modulations appear as the energy transitions from below to above $\Delta_0$ but disappear quickly leaving only two primary electronic structure elements of the pseudogap-energy electronic structure in $Z(\mathbf{q},E \cong \Delta_1)$. These are occur at $\mathbf{Q}_x$ =(1,0)$2\pi/a_0$ and $\mathbf{Q}_y$=(0,1) $2\pi/a_0$ which are the Bragg peaks representing the periodicity of the unit cell, and at $\mathbf{S}_x \equiv (\sim3/4,0)2\pi/a_0$, $\mathbf{S}_y \equiv (0,\sim3/4)2\pi/a_0$ which are due to the local breaking of lattice translation symmetry at the nanoscale. The doping evolution of $|\mathbf{S}_x|=|\mathbf{S}_y|$ (which is by definition that of $\mathbf{q}_5$* - see Fig. 5) as shown in Fig. 5d indicates that these incommensurate modulations are linked to the doping-dependence of the extinction point of the arc of Bogoliubov QPI.

Atomically resolved $\mathbf{r}$-space images of the static phenomena in $Z(\mathbf{r},E)$ show highly similar spatial patterns at all energies near $\Delta_1$ but with variations of intensity due to the $\Delta_1$-disorder (Fig. 3a). By changing to reduced energy variables $e(\mathbf{r}) = E/\Delta_1(\mathbf{r})$ and imaging $Z(\mathbf{r},e)$ it becomes clear that these modulations exhibit a strong maximum in intensity at $e = 1$. This is demonstrated directly in Figure 7 where the relative intensity of the modulations (all in same units and contrast scales) exhibits a strong maximum at $e$=1 [58]. Thus the pseudogap states of underdoped cuprates locally break translational symmetry, and reduce the expected 90°-rotational ($C_4$) symmetry of states within the unit cell to at least 180°-rotational ($C_2$) symmetry [57, 58, 59], and possibly to an even lower symmetry.

### 7.2  Universality of the Broken Symmetries of the $E \sim \Delta_1$ States

Theoretical concerns have been advanced about such spatial structuring of the cuprate pseudogap states including the possibility of spurious rotation symmetry breaking due to the dopant atoms [99]. To address such issues, we carried out a sequence of identical experiments on two radically different cuprates at the same $p$: strongly underdoped $Ca_{1.88}Na_{0.12}CuO_2Cl_2$ (Na-CCOC; $T_c \sim 21$ K) and $Bi_2Sr_2Dy_{0.2}Ca_{0.8}Cu_2O_{8+\delta}$ (Dy-Bi2212; $T_c \sim 45$ K). These materials have completely different crystallographic structures, chemical constituents, dopant-ion species, and inequivalent dopant-ion sites within the crystal-termination layers lying between the $CuO_2$ plane and the STM tip [57]. However images of the $E \sim \Delta_1$ pseudogap states for these two systems demonstrate a virtually indistinguishable electronic structure arrangements [57]. Obviously these symmetry breaking effects within every $CuO_2$ unit cell [57,68] cannot be governed by individual dopant ions because there is only a single such ion for every ~20 planar oxygen atoms in Dy-Bi2212. Moreover, the dopant ions occur at quite different locations (substitutional / interstitial respectively) in the unit cells of Na-CCOC and Dy-Bi2212. Thus, the virtually identical phenomena in images of the atomic-scale broken symmetries $E \sim \Delta_1$ pseudogap states in Na-CCOC and Dy-Bi2212 must occur due to the only common characteristic of these two radically different materials. Therefore $Z(\mathbf{r},e=1)$ images of the spatial structure of the cuprate pseudogap states [57, 58, 59, 68] should be ascribed to the intrinsic electronic structure of the $CuO_2$ plane.

*7.3    Imaging the Broken Spatial Symmetries of the Pseudogap $E \sim \Delta_1$ States*

To explore which spatial symmetries are actually broken by the cuprate pseudogap states, we use sub-unit-cell resolution $Z(\mathbf{r},e)$ imaging performed on multiple different underdoped $Bi_2Sr_2CaCu_2O_{8+\delta}$ samples with $T_c$'s between 20K and 55K. The necessary registry of the Cu sites in each $Z(\mathbf{r},e)$ is achieved by a picometer scale transformation which renders the topographic image $T(\mathbf{r})$ perfectly $a_0$-periodic; the same transformation is then applied to the simultaneously acquired $Z(\mathbf{r},e)$ to register all the electronic structure data to this ideal lattice. The topograph $T(\mathbf{r})$ is shown in Fig. 8a; the inset compares the Bragg peaks of its real (in-phase) Fourier components $\mathrm{Re}\,T(Q_x)$, $\mathrm{Re}\,T(Q_y)$ and demonstrates that $\mathrm{Re}\,T(Q_x)/\mathrm{Re}\,T(Q_y)=1$. Therefore $T(\mathbf{r})$ preserves the $C_4$ symmetry of the crystal lattice. In contrast, Fig. 8b shows that the $Z(\mathbf{r},e=1)$ determined simultaneously with Fig. 8a breaks various crystal symmetries [57-59]. The inset shows that since $\mathrm{Re}\,Z(Q_x,e=1)/\mathrm{Re}\,Z(Q_y,e=1) \neq 1$ the pseudogap states break $C_4$ symmetry on the average throughout Fig. 8b. We defined a normalized measure of intra-unit cell nematic ($C_2$) symmetry over the entire field of view (FOV) as a function of $e$:

$$O_N^Q(e) \equiv \frac{\mathrm{Re}\,Z(\mathbf{Q}_y,e) - \mathrm{Re}\,Z(\mathbf{Q}_x,e)}{\bar{Z}(e)} \qquad (7)$$

where $\vec{Z}(e)$ is the spatial average of $Z(\mathbf{r},e)$. The plot of $O_N^Q(e)$ in Fig. 8c shows that the magnitude of $O_N^Q(e)$ is low for $e \ll \Delta_0/\Delta_1$, begins to grow near $e \sim \Delta_0/\Delta_1$, and becomes well defined as $e \sim 1$ or $E \sim \Delta_1$. Thus the observed intra-unit-cell electronic symmetry breaking is specific to the pseudogap states.

To determine the source of these effects within the $CuO_2$ unit cell, we study $Z(\mathbf{r},e)$ with sub-unit-cell resolution. Fig. 8d shows the topographic image of a representative region from Fig. 8a; the locations of each Cu site **R** and of the two O atoms within its unit cell are indicated. Fig. 8e shows $Z(\mathbf{r},e)$ measured simultaneously with Fig. 8d with same Cu and O site labels. Next we define

$$O_N^R(e) = \sum_{\mathbf{R}} \frac{Z_x(\mathbf{R},e) - Z_y(\mathbf{R},e)}{\bar{Z}(e)N} \qquad (8)$$

where $Z_x(\mathbf{R},e)$ is the magnitude of $Z(\mathbf{r},e)$ at the O site $a_0/2$ along the x-axis from **R** while $Z_y(\mathbf{R},e)$ is the equivalent along the y-axis, and $N$ is the number of unit cells. This is the **r**-space measure of $C_2$ symmetry is equivalent of $O_N^Q(e)$ in Eqn. 7 but counting only O site contributions. Figure 8e contains the calculated value of $O_N^R(e)$ from the same FOV as Fig. 8a,b revealing the good agreement with $O_N^Q(e)$.

### 7.4  Separating $E\sim\Delta_1$ Broken Electronic Symmetry at *Q*=0 from that at *Q*=$S_x$, $S_y$

The smectic contributions to the $E\sim\Delta_1$ electronic structure can be examined by defining a measure analogous to Eqn. 7 of $C_4$ symmetry breaking, but now in the modulations with **S**$_x$, **S**$_y$ :

$$O_S^Q(e) = \frac{\operatorname{Re} Z(\mathbf{S_y},e) - \operatorname{Re} Z(\mathbf{S_x},e)}{\bar{Z}(e)} \qquad (9)$$

For all samples studied, the low values found for $|O_S^Q(e)|$ at low $e$ occur because these states are dispersive Bogoliubov quasiparticles [59, 53-56] and cannot be analyzed in term of any static electronic structure, smectic or otherwise. More importantly $|O_S^Q(e)|$ shows no tendency to become well established at the pseudogap or any other energy [68].

To visualize the separate broken symmetries in the $E\sim\Delta_1$ electronic structure, we consider $Z(\mathbf{q},e=1)$ in Fig. 9a; this is the Fourier-space representation of electronic structure of the $E\sim\Delta_1$ states. Taking into account only the Bragg peaks at $\mathbf{Q}_x$, $\mathbf{Q}_y$ (red circles/arrows in Fig. 9a) the $C_4$ symmetry breaking of **Q**=0 intra-unit-cell electronic structure is revealed as shown schematically in Fig. 9b. By contrast, if one focuses upon the incommensurate modulations $\mathbf{S}_x$, $\mathbf{S}_y$ (blue circles/arrows in Fig 9a), we find a disordered electronic structure with incommensurate modulations which break both $C_2$ and translational symmetry locally as shown schematically in Fig. 9c. Although these two types of electronic phenomena represent clearly distinct broken symmetries, SI-STM reveals that they coexist in the $E\sim\Delta_1$ pseudogap electronic structure of underdoped cuprates [68].

*7.5    Summary*

When $Z(\mathbf{r},E)$ images of the intra-unit-cell electronic structure in underdoped $Bi_2Sr_2CaCu_2O_{8+\delta}$ are analyzed using two independent techniques, compelling evidence for intra-unit-cell (or $\mathbf{Q}=0$) electronic symmetry breaking is detected specifically of the states at the $E\sim\Delta_1$ pseudogap energy. Moreover, this intra-unit-cell symmetry breaking coexists with finite $\mathbf{Q}=\mathbf{S_x}, \mathbf{S_y}$ smectic electronic modulations, but they can be analyzed separately by using Fourier filtration techniques. The wavevector of smectic electronic modulations is controlled by the point in $\mathbf{k}$-space where the Bogoliubov interference signature disappears when the arc supporting delocalized Cooper pairing approaches the lines between $\mathbf{k} =\pm(0,\pi/a_0)$ and $\mathbf{k} =\pm(\pi/a_0,0)$ (see Fig. 5b,d). This appears to indicate that the $\mathbf{Q}=\mathbf{S_x}, \mathbf{S_y}$ smectic effects are dominated by the same $\mathbf{k}$-space phenomena which restrict the regions of Cooper pairing [58].

## 8    Overview, Conclusions and Future

*8.1    Bipartite Electronic Structure of Underdoped Cuprates derived from SI-STM*

A clearer picture of the fundamentally bipartite electronic structure of strongly underdoped cuprates approaching the Mott insulator emerges from these SI-STM studies. This is summarized in Fig. 10. In the dSC phase (Fig. 10a,b,c) the Bogoliubov QPI signature of delocalized Cooper pairs (Section 5 and Fig. 10c) exists upon the arc in $\mathbf{k}$-space labeled by region II in Fig. 10a. The Bogoliubov QPI disappears near the lines connecting $\mathbf{k}=(0,\pm\pi/a_0)$ to $\mathbf{k}=(\pm\pi/a_0,0)$ - thus defining a $\mathbf{k}$-space arc which supports the delocalized Cooper pairing. This arc shrinks rapidly towards the $\mathbf{k}=(\pm\pi/2a_0,\pm\pi/2a_0)$ points with falling hole-density in a fashion which could satisfy Luttinger's theorem if it were actually a hole-pocket bounded from behind by the $\mathbf{k}=\pm(\pi/a_0,0)$ - $\mathbf{k}=\pm (0,\pi/a_0)$ lines.  The $E\sim\Delta_1$ pseudogap excitations (Section 7) are labeled by region I in Fig. 10a and exhibit a radically different $\mathbf{r}$-space phenomenology locally breaking the expected $C_4$ symmetry of electronic structure at least down to $C_2$ and possibly to an even lower symmetry, within each $CuO_2$ unit cell (Fig. 10b). These $\mathbf{Q}=0$ broken electronic symmetry states coexist with finite $\mathbf{Q}=\mathbf{S_x}, \mathbf{S_y}$ modulations which break translational and rotational symmetry very locally. In the PG phase (Fig. 10d,e,f), the Bogoliubov QPI signature (Section 6 and Fig. 10f) exists upon a smaller part of the same arc in $\mathbf{k}$-space as it did in the dSC phase. This is labeled as region II in Fig. 10d. Here, however, since the ungapped Fermi arc (region III) predominates, the gapped region supporting d-wave QPI has shrunk into a narrow sliver near a line connecting $\mathbf{k}=(\pi/a_0,0)$ and $\mathbf{k}=(0,\pi/a_0,)$ (Fig. 10d). The $E\sim\Delta_1$ excitations in the PG phase, (Section 7) are again labeled by region I in Fig. 10d and exhibit $\mathbf{Q}=0$ and $\mathbf{Q}=\mathbf{S_x}, \mathbf{S_y}$ broken electronic symmetries indistinguishable from those in the dSC phase (Fig. 10e).

*8.2    Microscopic Mechanism of Intra-unit-cell Electronic Symmetry Breaking*

The microscopic source of the intra-unit-cell ($\mathbf{Q}=0$) electronic symmetry breaking in the $E\sim\Delta_1$ states (Fig. 8) is unknown at present. One important point to consider is the relationship between ARPES, elastic neutron scattering (NS) and SI-STM studies of broken electronic symmetries of the PG phase. ARPES reveals spontaneous dichroism of antinodal states [100] which break $C_4$ symmetry because the opposite sign of the effect occurs at $\mathbf{k}=(\pi/a_0,0)$ and $\mathbf{k}=(0,\pi/a_0)$ The $\mathbf{Q}=0$ magnetic order detected by NS at the Bragg peak [101, 102] consists of intra-unit cell, apparently antiferromagnetic and $C_4$-breaking states in both $YBa_2Cu_3O_{6+x}$ and $HgBa_2CuO_{4+\delta}$. The SI-STM studies also reveal intra-unit cell, $C_4$-breaking states at the pseudogap energy and show that these effects are associated primarily with electronic inequivalence at the two O sites within the $CuO_2$ unit cell (Section 7). With such commonality between the results from such disparate techniques, it is not implausible they are detecting different characteristics of the same broken symmetry states. If so, an immediate consequence of the existence of the $\mathbf{Q}=0$ electronic/magnetic structures within the $CuO_2$ unit cell, would be that an effective model defined purely on the copper lattice (such as the t-J type of model) will be unable to capture the physics of underdoped cuprates.

*8.3    Relationship between the two Broken Electronic Symmetries and the Superconductivity*

Both nematic and smectic broken symmetries have been reported in the electronic structure of different cuprate compounds [103-106]. A spin/charge smectic broken symmetry phase (stripes) exists in $La_{2-x-y}Nd_ySr_xCuO_4$ and $La_{2-x}Ba_xCuO_4$ when $p\sim0.125$. Nematic broken symmetry has been reported in underdoped $YBa_2Cu_3O_{6+\delta}$ [101], underdoped $Bi_2Sr_2CaCu_2O_{8+\delta}$ [68, 100] and underdoped $HgBa_2CuO_{4+x}$ [102]. To understand how both these distinct broken symmetry states can coexist, and to determine the form of their interactions, will be important in unraveling the mystery of the cuprate phase diagram. That equivalent broken symmetries appear to coexist at the nanoscale in the electronic structure of $Bi_2Sr_2CaCu_2O_{8+\delta}$ ([68] and Fig. 9) represents an important new opportunity to understand their interactions. Should that be possible, the next challenge for SI-STM would be to demonstrate directly the relationship between the superconductivity and the broken symmetries of the $E\sim\Delta_1$ pseudogap states with the (ambitious) view towards a complete Ginzburg-landau understanding the cuprate phase diagram.

*8.4    Electronic Structure of the Cuprate Pseudogap Phase*

Among the explanations for the PG phase is that it is a spin liquid created by hole-doping an antiferromagnetic Mott insulator, or that it is a d-wave superconductor without phase-coherence, or that it is an electronic ordered phase with additional broken symmetries. SI-STM reveals that the basic particle-hole symmetric, dispersive, octet phenomenology is consistent with theoretical predictions for the QPI characteristics of a phase incoherent d-wave superconductor (Fig. 10f). Further, since all the $\mathbf{q}_i(E)$ $i=1,...7$ disperse internally consistently with the octet-model, they cannot represent the signature of any static ordered state of fixed wavevector $\mathbf{Q}^*$. Thus the low energy $E<\Delta_0$ electronic structure of the PG phase (which is what is probed by transport and thermodynamics) is indeed consistent with expectations for a phase-incoherent d-wave superconductor. Nevertheless, the high energy electronic states at the pseudogap energy scale $E\sim\Delta_1$ exhibit strongly broken symmetries including intra-unit-cell symmetry breaking and finite $\mathbf{Q}$ smectic modulations (Fig. 9). Finally, the truncated arc of Bogoliubov QPI seen below $T_c$, which remains unchanged in the PG phase except for the appearance of an ungapped portion, appears not-inconsistent with the phenomenological models proposed for a spin liquid (see below). Thus the characteristics of the PG phase determined by SI-STM contain some elements of all three theoretical approaches to the electronic structure of hole-doped $CuO_2$ approaching the Mott insulator.

*8.5    Fundamental Electronic Structure of the Hole-doped $CuO_2$ Mott Insulator from SI-STM*

The overall electronic structure of underdoped cuprates as derived from SI-STM studies (Fig. 10) motivates a number of questions. Why does the Bogoliubov QPI signature of delocalized Cooper pairs disappear [58] near the $\mathbf{k}=(0,\pm\pi/a_0) - \mathbf{k}=(\pm\pi/a_0,0)$ connecting lines? And why do the pseudogap states $E\sim\Delta_1$ exhibit such dramatically different symmetries [57, 68] to the coexisting Bogoliubov quasiparticles at $E<\Delta_0$? One reason could be that the $\mathbf{r}$-space electronic structure has undergone a $\sqrt{2}\times\sqrt{2}$ reconstruction due to the appearance of a coexisting long-range ordered state. The arcs supporting Cooper pairing would then represent one side of a hole-pocket within a reduced Brillouin zone. But neither antiferromagnetism nor other long-range ordered electronic phases [23, 24] necessary for such a reconstruction has yet been detected in $Bi_2Sr_2CaCu_2O_{8+\delta}$. A related explanation could be inelastic scattering of the quasiparticles by spin fluctuations [107, 108] at $\mathbf{Q}=(\pi/a_0,\pi/a_0)$ or by fluctuations of other ordered states which would exhibit a $\sqrt{2}\times\sqrt{2}$ reconstruction if stabilized. Neither of these approaches explains the broken spatial symmetries of the $E\sim\Delta_1$ pseudogap states, however. Another type of explanation could be a spin-charge stripe glass [57] coexisting with superconductivity [109-112]. This could explain the loss of translational symmetry and the $C_4$ breaking within the $E\sim\Delta_1$ pseudogap states, and perhaps the disappearance of quasiparticle interference along the $\mathbf{k}=(0,\pm\pi/a_0) - \mathbf{k}=(\pm\pi/a_0,0)$ lines [90,113], but it does not (yet) explain the intra-unit-cell $C_4$ breaking in electronic symmetry. Yet another proposal, that orbital charge currents exist within each $CuO_2$ unit-cell [22], receives support from NS experiments [101, 102] and may provide an explanation for the intra-unit-cell electronic symmetry breaking discussed here (although reasons why such an orbiting current could be detected by SI-STM are unknown). But it does not (yet) explain the finite **Q** smectic modulations or the disappearance of Bogoliubov QPI near $\mathbf{k}=(0,\pm\pi/a_0) - \mathbf{k}=(\pm\pi/a_0,0)$ lines. A final possibility, which is revealed by the fact that the Luttinger theorem can be satisfied by using the region bounded the Bogoliubov QPI arcs and the $\mathbf{k}=(0,\pm\pi/a_0) - \mathbf{k}=(\pm\pi/a_0,0)$ lines [58], is that many of the effects summarized in Fig. 10 are properties of a hole-doped spin liquid [15]. This approach might explain (at least phenomenologically) the Bogoliubov arc termination as where the Green's-function poles turn to zeros along the $\mathbf{k}=(0,\pm\pi/a_0) - \mathbf{k}=(\pm\pi/a_0,0)$ lines [15, 58], how the Luttinger theorem can be satisfied given the exotic **k**-space structure observed [15, 58], and possibly the cause of smectic finite-**Q** non-dispersive modulations [114]. However it does not (yet) appear to explain **Q**=0 intra-unit-cell electronic symmetry breaking.

When the electronic structure of underdoped cuprates is examined with high resolution in both **r**-space and **k**-space using SI-STM, a highly complex phenomenology is revealed. As is often the case, if one focuses on a single element within such a ramified phenomenology, there are several theoretical models available to explain it. One hopes, however, that the eventual overarching theory of cuprate high temperature superconductivity will explain the complete phenomenology in a unified fashion - as Bardeen-Cooper-Schrieffer theory did for conventional superconductors. In this review we attempt to contribute to such an aspiration by summarizing what we think are the most important elements of cuprate electronic structure phenomenology revealed by a decade of SI-STM studies. The key questions emerging from this effort are whether the observed broken symmetries (and/or perhaps others yet to be discovered) are responsible for the opening of the pseudogap and, if so, how these exotic broken symmetry states interact with the superconducting components of the $CuO_2$ electronic structure.

**Acknowledgements**

We acknowledge and thank all our collaborators: J.W. Alldredge, I. Firmo, M.H. Hamidian, T. Hanaguri, P. J. Hirschfeld, J.E. Hoffman, E.W. Hudson, Chung Koo Kim, Y. Kohsaka, K.M. Lang, C. Lupien, Jhinhwan Lee, Jinho Lee, V. Madhavan, K. McElroy, J. Orenstein, S.H. Pan, R. Simmonds, J. Slezak, J. Sethna, H. Takagi, C. Taylor, P. Wahl, & M. Wang. Preparation of this review was supported by the Center for Emergent Superconductivity, an Energy Frontier Research Center funded by the U.S. Department of Energy, Office of Basic Energy Sciences under Award Number DE-2009-BNL-PM015.

**Figure Captions**

**Figure 1. a**, Schematic copper-oxide phase diagram. Here $T_C$ is the critical temperature circumscribing a 'dome' of superconductivity, $T_\phi$ is the maximum temperature at which superconducting phase fluctuations are detectable within the pseudogap phase, and $T^*$ is the approximate temperature at which the pseudogap phenomenology first appears. **b,** The two classes of electronic excitations in cuprates. The separation between the energy scales associated with excitations of the superconducting state (dSC, denoted by $\Delta_0$) and those of the pseudogap state (PG, denoted by $\Delta_1$) increases as *p* decreases (reproduced from [7]). The different symbols correspond to the use of different experimental techniques.

**Figure 2. a,** Fourier transform of the conductance ratio map $Z(\mathbf{r}, E)$ at a representative energy below $\Delta_0$ for $T_C$ = 45K $Bi_2Sr_2Dy_{0.2}Ca_{0.8}Cu_2O_{8+\delta}$, which only exhibits the patterns characteristic of homogenous *d*-wave superconducting quasiparticle interference. **b,** Evolution of the spatially averaged tunneling spectra of $Bi_2Sr_2CaCu_2O_{8+\delta}$ with diminishing *p*, here characterized by $T_C(p)$. The energies $\Delta_1(p)$ (blue dashed line) are easily detected as the pseudogap edge while the energies $\Delta_0(p)$ (red dashed line) are more subtle but can be identified by the correspondence of the "kink" energy with the extinction energy of Bogoliubov quasiparticles, following the procedures in refs. [53,59] . **c,** Laplacian of the conductance ratio map $Z(\mathbf{r})$ at the pseudogap energy $E = \Delta_1$, emphasizing the local symmetry breaking of these electronic states for strongly underdoped $Ca_{1.88}Na_{0.12}CuO_2Cl_2$.

**Figure 3. a,** Map of the local energy scale $\Delta_1(\mathbf{r})$ from a 49nm field of view (corresponding to ~16,000 $CuO_2$ plaquettes) measured on a sample with $T_C$ = 74K. Average gap magnitude $\Delta_1$ is at the top, together

with the values of N, the total number of dopant impurity states (shown as white circles) detected in the local spectra. **b,** The average tunneling spectrum, $g(E)$, associated with each gap value in the field of view in **a**. The arrows locate the "kinks" whose energy is $\Delta_0$ **c,** Histograms of equivalent $\Delta_1(\mathbf{r})$ maps from samples with $p$ = 0.08, 0.10, 0.14, 0.17, 0.19, and 0.22 normalized to the average $\Delta_1$ in each map. Obviously, these distributions are statistically highly similar. **d,** The doping dependence the average $\Delta_1$ (blue circles), average $\Delta_0$ (red circles) and average antinodal scattering rate $\Gamma_2^*$ (black squares), each set interconnected by dashed guides to the eye. The higher-scale $\Delta_1$ evolves along the pseudogap line whereas the lower-scale $\Delta_0$ represents segregation in energy between homogeneous and heterogeneous electronic structure.

**Figure 4. a,** The expected wavevectors of quasiparticle interference patterns in a superconductor with electronic band structure like that of $Bi_2Sr_2CaCu_2O_{8+x}$. Solid lines indicate the **k**-space locations of several banana-shaped quasiparticle contours of constant energy as they increase in size with increasing energy. As an example, at a specific energy, the octet of regions of high JDOS are shown as red circles. The seven primary scattering **q**-vectors interconnecting elements of the octet are shown in blue. **b,** A plot of the superconducting energy gap $\Delta(\theta_k)$ determined from octet model inversion of quasiparticle interference measurements, shown as open circles [55]. These were extracted using the measured position of scattering vectors $\mathbf{q}_1$ through $\mathbf{q}_7$. The solid line is a fit to the data. The mean value of $\Delta_1$ for this overdoped $T_C$ = 86K sample was 39 meV. **c,** The magnitude of various extracted QPI vectors, plotted as a function of energy. Whereas the expected energy dispersion of the octet vectors $\mathbf{q}_i(E)$ is apparent for $|E|$ < 32mV, the peaks which avoid extinction ($\mathbf{q}_1^*$ and $\mathbf{q}_5^*$) always become non-dispersive above $\Delta_0$ (vertical grey line). **d,** Locus of the Bogoliubov band minimum $\mathbf{k}_B(E)$ found from extracted QPI peak locations $\mathbf{q}_i(E)$, in five independent $Bi_2Sr_2CaCu_2O_{8+x}$ samples with decreasing hole density. Fits to quarter-circles are shown and, as $p$ decreases, these curves enclose a progressively smaller area. The BQP interference patterns disappear near the perimeter of a **k**-space region bounded by the lines joining $\mathbf{k} = (0, \pm\pi/a_0)$ and $\mathbf{k} = (\pm\pi/a_0, 0)$. The spectral weights of $\mathbf{q}_2$, $\mathbf{q}_3$, $\mathbf{q}_6$ and $\mathbf{q}_7$ vanish at the same place (dashed line; see also ref. [59]). Filled symbols in the inset represent the hole count $p = 1 - n$ derived using the simple Luttinger theorem, with the fits to a large, hole-like Fermi surface indicated schematically here in grey. Open symbols in the inset are the hole counts calculated using the area enclosed by the Bogoliubov arc and the lines joining $\mathbf{k} = (0, \pm\pi/a_0)$ and $\mathbf{k} = (\pm\pi/a_0, 0)$, and are indicated schematically here in blue.

**Figure 5. a,** Fourier transform of the conductance ratio $Z(\mathbf{q}, E=48\text{meV})$ at a representative energy between $\Delta_0$ and $\Delta_1$ for underdoped $T_C$=74K $Bi_2Sr_2CaCu_2O_{8+x}$. The red line schematically indicates the source of the data in **c**. The arrows label the location of the wavevectors $\mathbf{q}_1^*$, $\mathbf{q}_5^*$, $S_x$, and $Q_x$ described in the text. **b,** Schematic diagram of the Brillouin zone illustrating the relationship of non-dispersive $\mathbf{q}_1^*$ and $\mathbf{q}_5^*$ to the ends of the Bogoliubov arc. **c,** Doping dependence of line-cuts of $Z(\mathbf{q}, E=48\text{meV})$ extracted along the Cu-O bond direction $\mathbf{Q}_x$. The vertical dashed lines demonstrate that the non-dispersive **q**-vectors at energies between $\Delta_0$ and $\Delta_1$ are not commensurate harmonics of a $4a_0$ periodic modulation, but instead evolve in a fashion directly related to the extinction point of the Fermi arc. The data in **c**. have been normalized to the peak amplitude of $\mathbf{q}_5^*$ and offset vertically for clarity. **d,** $\mathbf{q}_1^*$, $\mathbf{q}_5^*$, and their sum $\mathbf{q}_1^* + \mathbf{q}_5^*$ as a function of $p$ demonstrating that individually these modulations evolve with doping while their sum does not change and is equal to the reciprocal lattice vector defining the first Brillouin zone.

**Figure 6.** (**a** to **x**) Differential conductance maps $g(\mathbf{r},E)$ were obtained on the same sample in an atomically resolved and registered FOV > 45 × 45 nm$^2$ at six temperatures. Each panel shown is the Fourier transform $Z(\mathbf{q},E)$ of $Z(\mathbf{r},E) \equiv g(\mathbf{r},+E)/g(\mathbf{r},-E)$ for a given energy and temperature. The QPI signals evolve dispersively with energy along the horizontal energy axis. The temperature dependence of QPI for a given energy evolves along the vertical axis. The octet-model set of QPI wave vectors is observed for every $E$ and $T$ as seen, for example, by comparing (**a**) and (**u**), each of which has the labeled octet vectors. Within the basic octet QPI phenomenology, there is no particular indication in these data of where the superconducting transition $T_C$, as determined by resistance measurements, occurs.

**Figure 7**. A series of images displaying the real space conductance ratio $Z(\mathbf{r},e)$ as a function of energy rescaled to the local pseudogap value, $e = E/\Delta_1(\mathbf{r})$. Each pixel location was rescaled independently of the others. The common color scale for all panels illustrates that the broken electronic symmetry patterns appear strongest in $Z$ exactly at $E = \Delta_1(\mathbf{r})$, the local pseudogap energy.

**Figure 8. a,** Topographic image $T(\mathbf{r})$ of the $Bi_2Sr_2CaCu_2O_{8+\delta}$ surface. The inset shows that the real part of its Fourier transform Re $T(\mathbf{q})$ does not break $C_4$ symmetry at its Bragg points because plots of $T(\mathbf{q})$ show its values to be indistinguishable at $\mathbf{Q}_x = (1, 0)2\pi/a_0$ and $\mathbf{Q}_y = (0, 1)2\pi/a_0$. Importantly, this means that neither the crystal nor the tip used to image it (and its $Z(\mathbf{r}, E)$ simultaneously) exhibits $C_2$ symmetry. **b,** The $Z(\mathbf{r}, e = 1)$ image measured simultaneously with $T(\mathbf{r})$ in **a**. The inset shows that the Fourier transform $Z(\mathbf{q}, e = 1)$ does break $C_4$ symmetry at its Bragg points because Re $Z(\mathbf{Q}_x, e\sim1) \neq$ Re $Z(\mathbf{Q}_y, e\sim1)$ . This means that, on average throughout the FOV of **a** and **b**, the modulations of $Z(\mathbf{r}, E < \Delta_1)$ that are periodic with the lattice have different intensities along the $x$ axis and along the $y$ axis. **c,** The value of $O_N^Q(e)$ defined in Eqn. 7 computed from $Z(\mathbf{r}, e)$ data measured in the same FOV as **a** and **b**. Its magnitude is low for all $E < \Delta_0$ and then rises rapidly to become well established near $e < 1$ or $E < \Delta_1$. Thus the quantitative measure of intra-unit-cell electronic nematicity reveals that the pseudogap states in this FOV of a strongly underdoped $Bi_2Sr_2CaCu_2O_{8+\delta}$ sample break the expected $C_4$ symmetry of $CuO_2$ electronic structure. **d,** Topographic image $T(\mathbf{r})$ from the region identified by a small white box in **a**. It is labeled with the locations of the Cu atom plus both the O atoms within each $CuO_2$ unit cell (labels shown in the inset). Overlaid is the location and orientation of a Cu and four surrounding O atoms. **e,** The simultaneous $Z(\mathbf{r}, e = 1)$ image in the same

FOV as **d** (the region identified by small white box in b) showing the same Cu and O site labels within each unit cell (see inset). Thus the physical locations at which the nematic measure $O_N^R(e)$ of Eqn. 8 is evaluated are labeled by the dashes. Overlaid is the location and orientation of a Cu atom and four surrounding O atoms. **f,** The value of $O_N^R(e)$ computed from $Z(\mathbf{r}, e)$ data measured in the same FOV as **a** and **b**. As in **c**, its magnitude is low for all $E < \Delta_0$ and then rises rapidly to become well established at $e \sim 1$ or $E \sim \Delta_1$.

**Figure 9. a** The Fourier transform $Z(\mathbf{q},e=1)$ of a typical image $Z(\mathbf{r},e=1)$ of the spatial structure of the pseudogap states in underdoped $Bi_2Sr_2CaCu_2O_{8+\delta}$. The Bragg peaks are identified by red circles and $\mathbf{Q}_x$, $\mathbf{Q}_y$ labels. The wavevectors of the smectic modulations in electron structure are identified by blue circles and $\mathbf{S}_x$, $\mathbf{S}_y$ labels. **b** Schematic depiction of how the spatial information in the inequivalent Bragg peaks $\mathbf{Q}_x$, $\mathbf{Q}_y$ alone could reveal intra unit-cell $C_2$-symmetric electronic structure. **c** Schematic depiction of how the spatial information in the $\mathbf{S}_x$, $\mathbf{S}_y$ wavevectors alone can reveal the disordered breaking of both rotational and translational symmetry in electronic structure.

**Figure 10. a** A schematic representation of the electronic structure in one quarter of the Brillouin zone at lowest temperatures in the dSC phase. The region marked II in front of the line joining $\mathbf{k}=(\pi/a_0,0)$ and $\mathbf{k}=(0,\pi/a_0)$ is the locus of the Bogoliubov QPI signature of delocalized Cooper pairs. **b** An example of the broken spatial symmetries which are concentrated upon the pseudogap energy $E\sim\Delta_1$ as measured at lowest temperatures. **c** An example of the characteristic Bogoliubov QPI signature of sixteen pairs of interference wavevectors, all dispersive and internally consistent with the octet model as well as particle-hole symmetric $\mathbf{q}_i(+E)=\mathbf{q}_i(-E)$, here measured at lowest temperatures. **d** A schematic representation of the electronic structure in one quarter of the Brillouin zone at $T\sim 1.5\ T_c$ in the PG phase. The region marked III is the Fermi arc, which is seen in QPI studies as a set of interference wavevectors $\mathbf{q}_i(E=0)$ which indicate that there is no gap-node at $E=0$. Region II in front of the line joining $\mathbf{k}=(\pi/a_0,0)$ and $\mathbf{k}=(0,\pi/a_0)$ is the locus of the phase incoherent Bogoliubov QPI signature. Here all 16 pairs of wavevectors of the octet model are detected and found to be dispersive. Thus although the sample is not a long-range phase coherent superconductor, it does give clear QPI signatures of d-wave Cooper pairing. **e** An example of the broken spatial symmetries which are concentrated upon pseudogap energy $E\sim\Delta_1$ as measured in the PG phase; they are indistinguishable from measurements at $T\sim 0$. **f** An example of the characteristic Bogoliubov QPI signature of sixteen pairs of interference wavevectors, all dispersive and internally consistent with the octet model as well as particle-hole symmetric $\mathbf{q}_i(+E)=\mathbf{q}_i(-E)$, here measured at $T\sim 1.5T_c$.

Figure 1

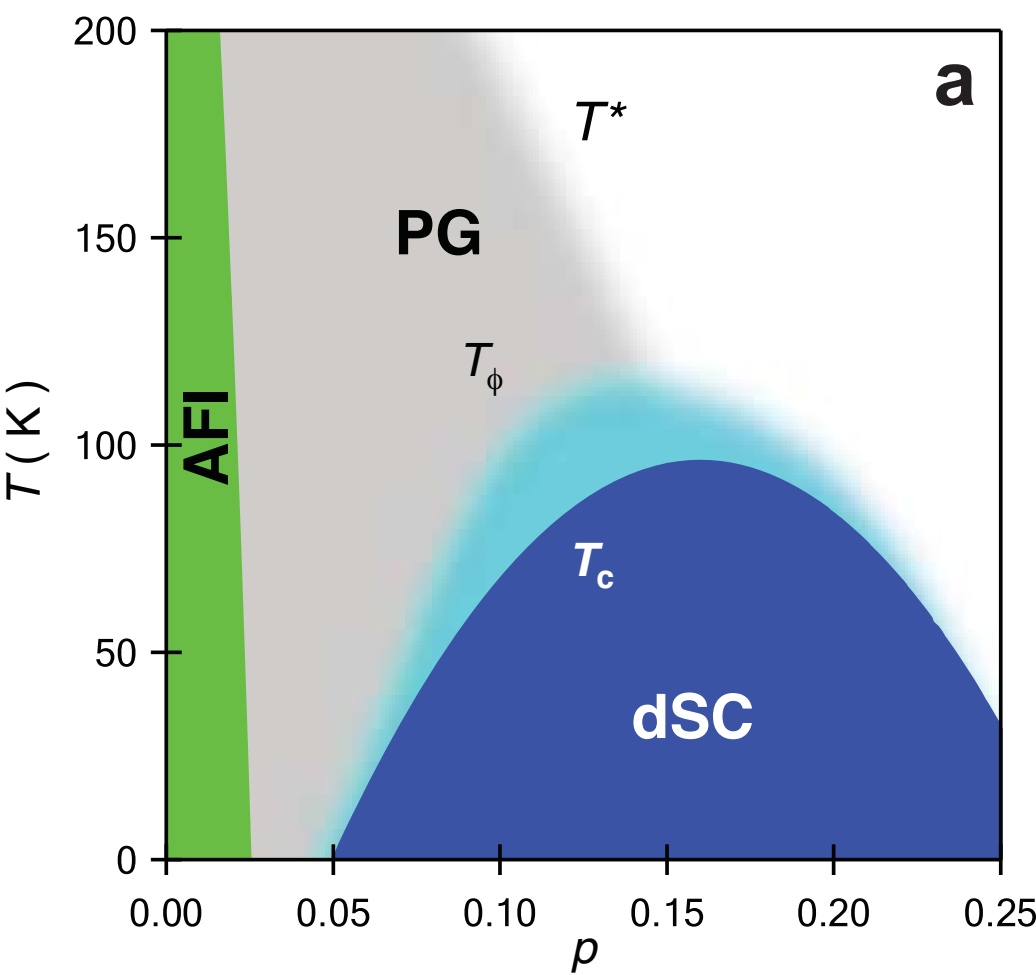

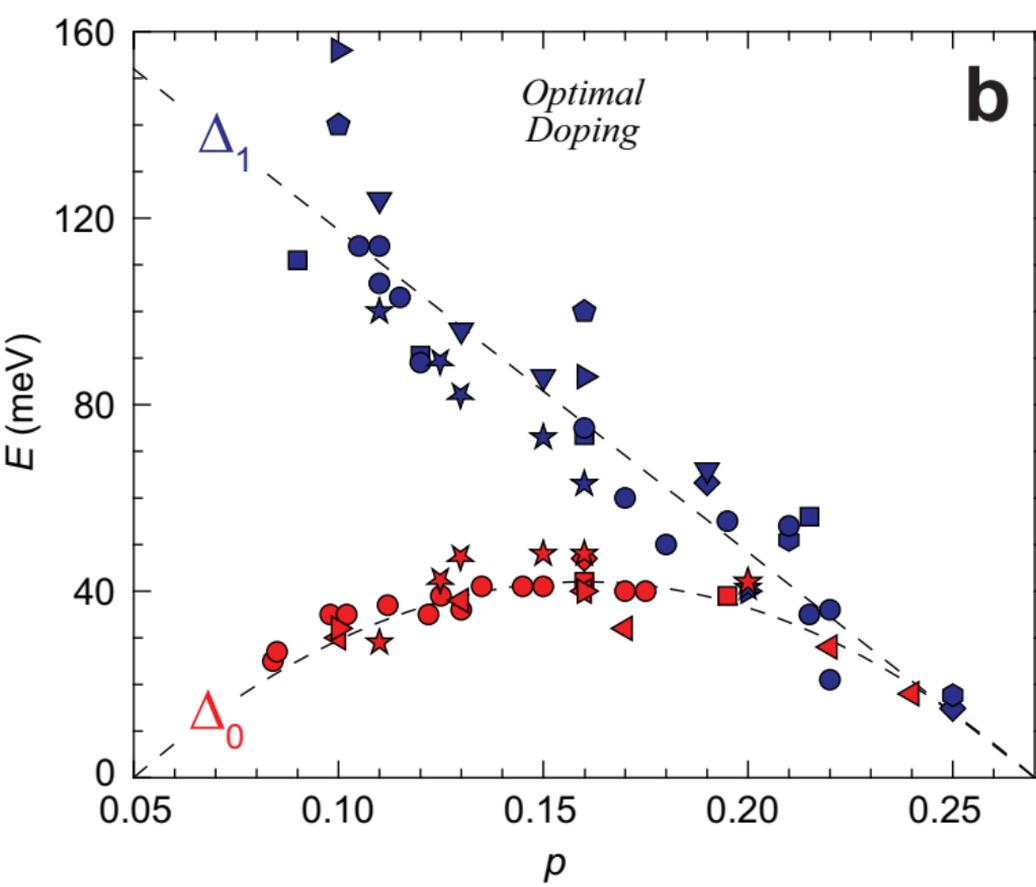

Figure 2

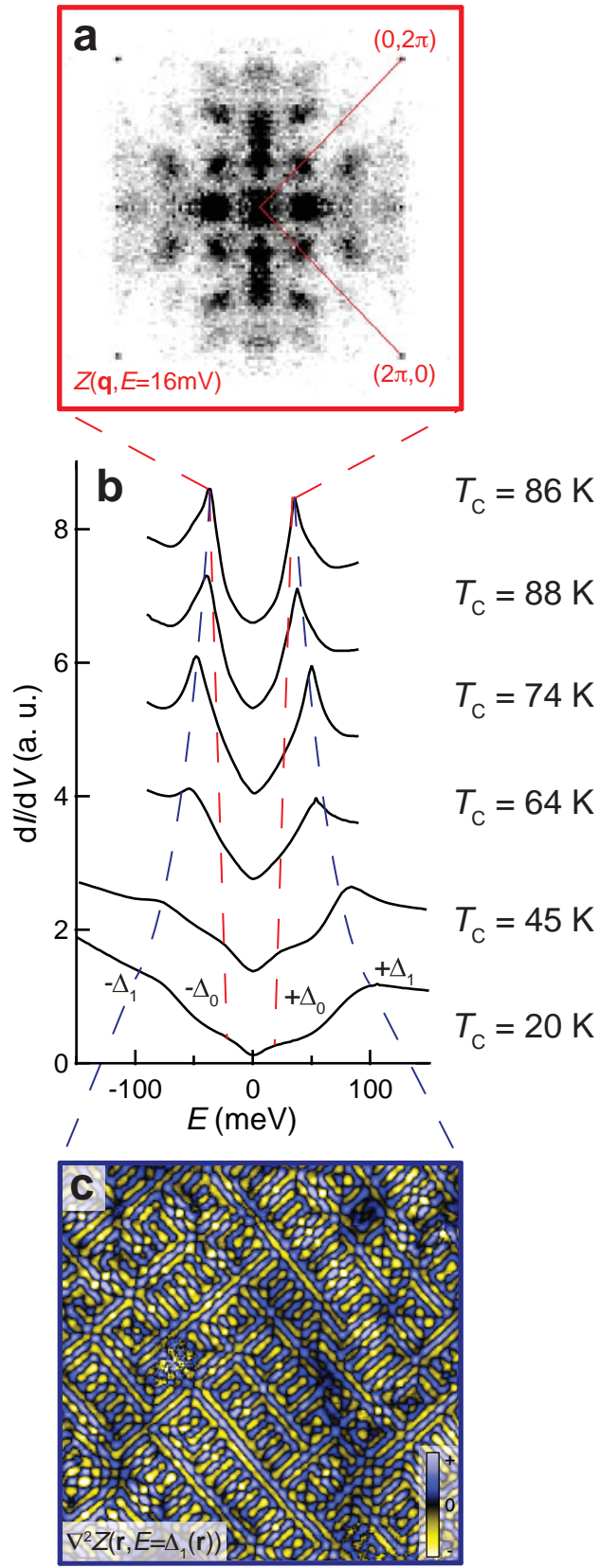

Figure 3

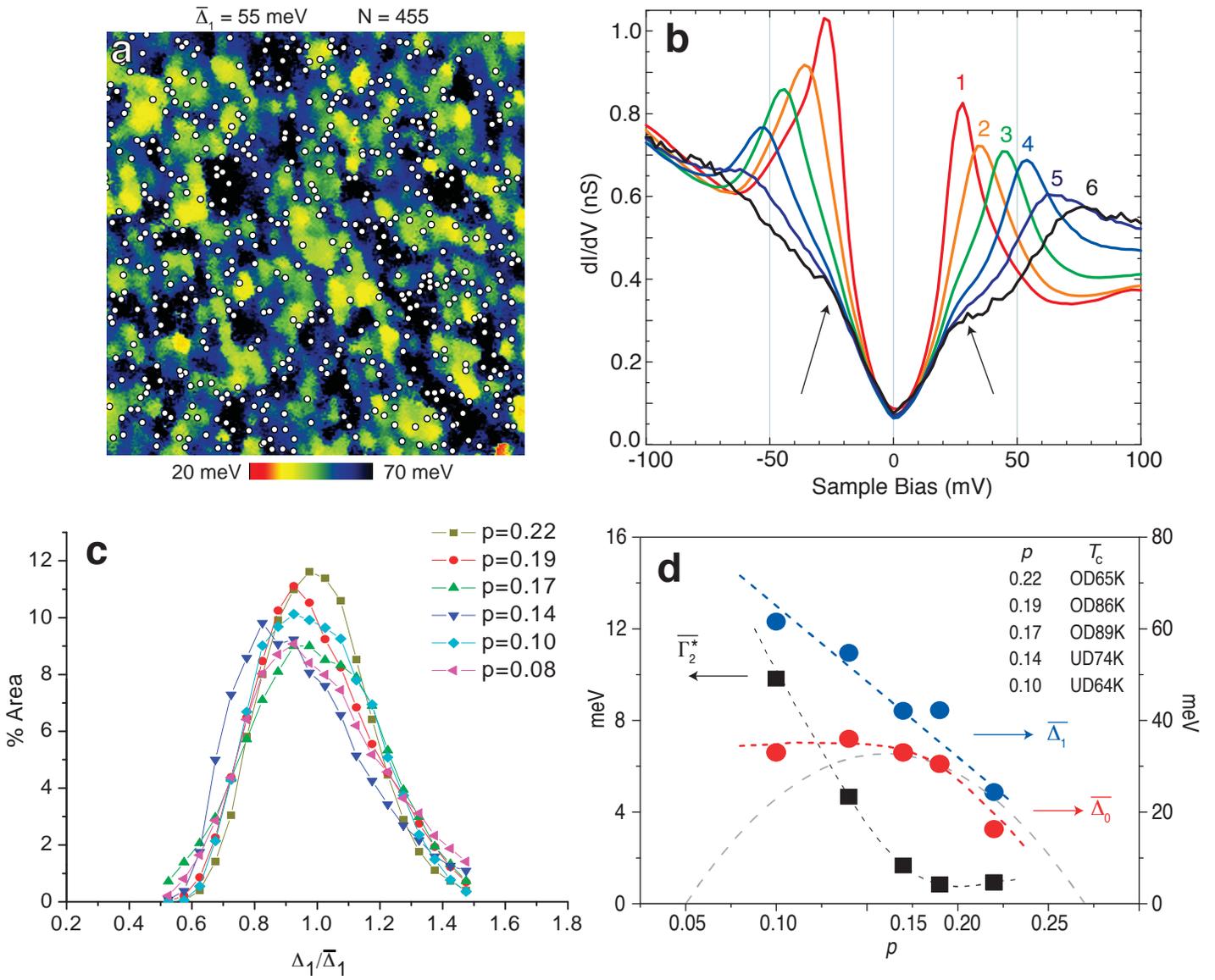

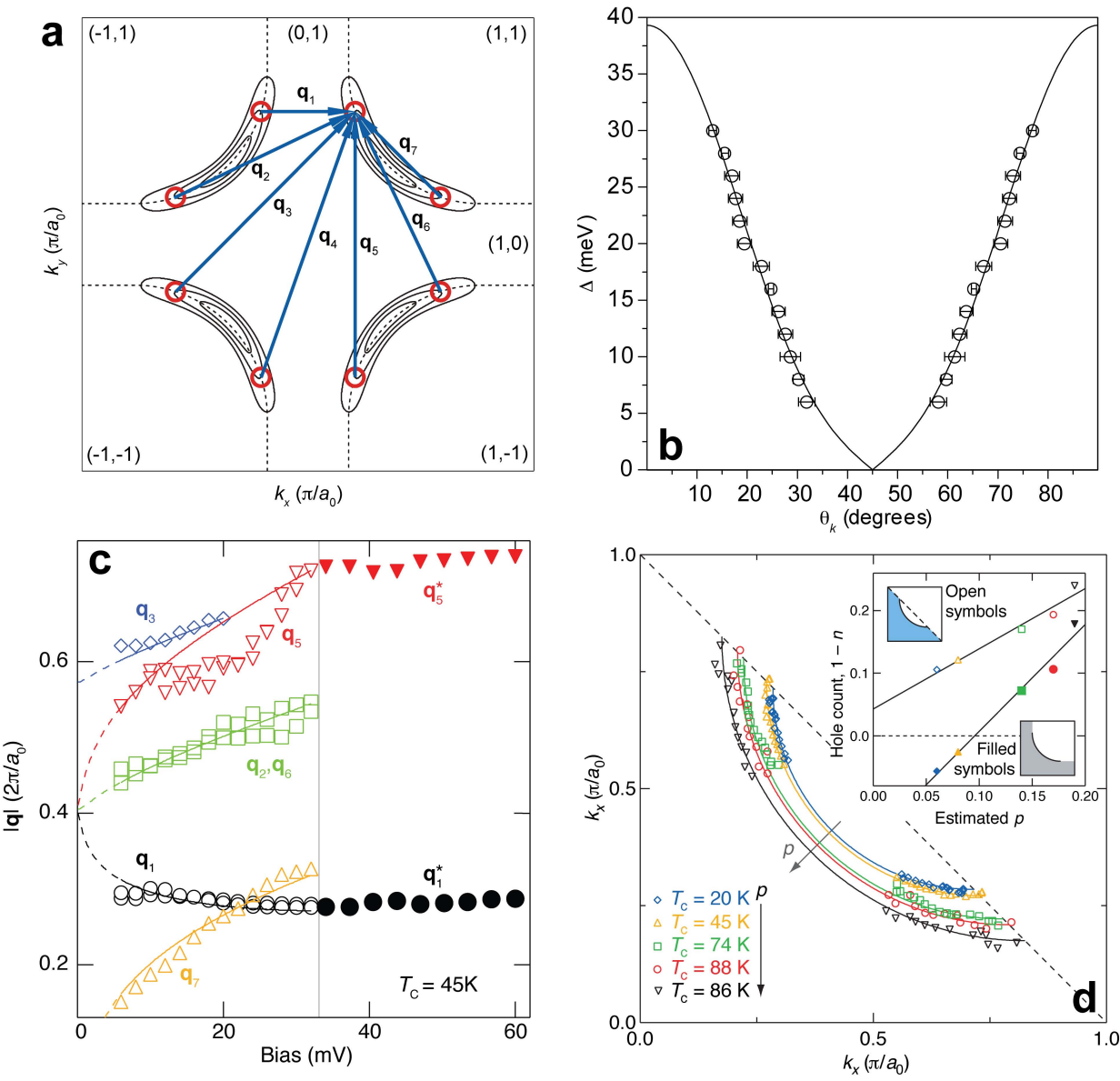

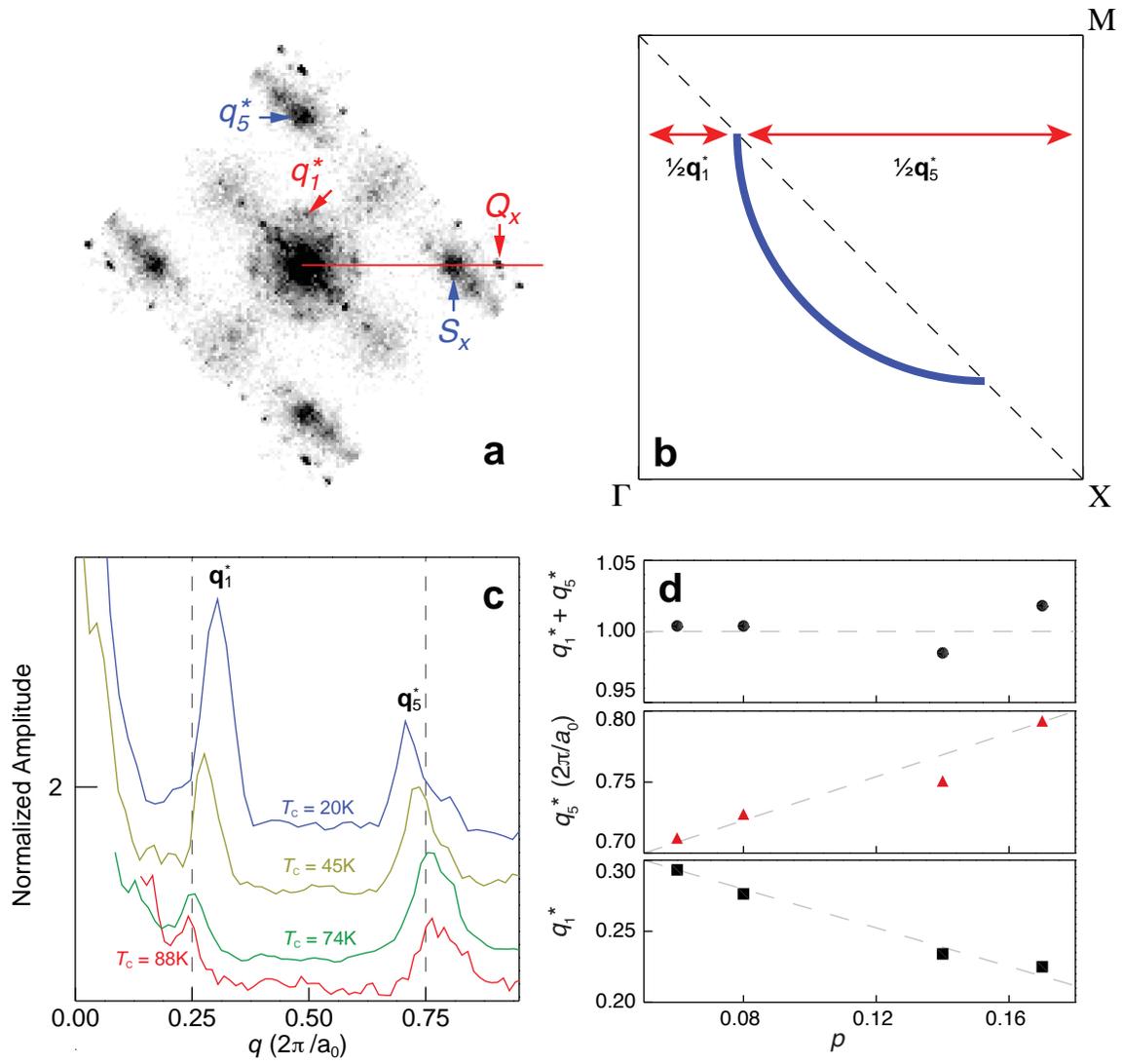

Figure 6

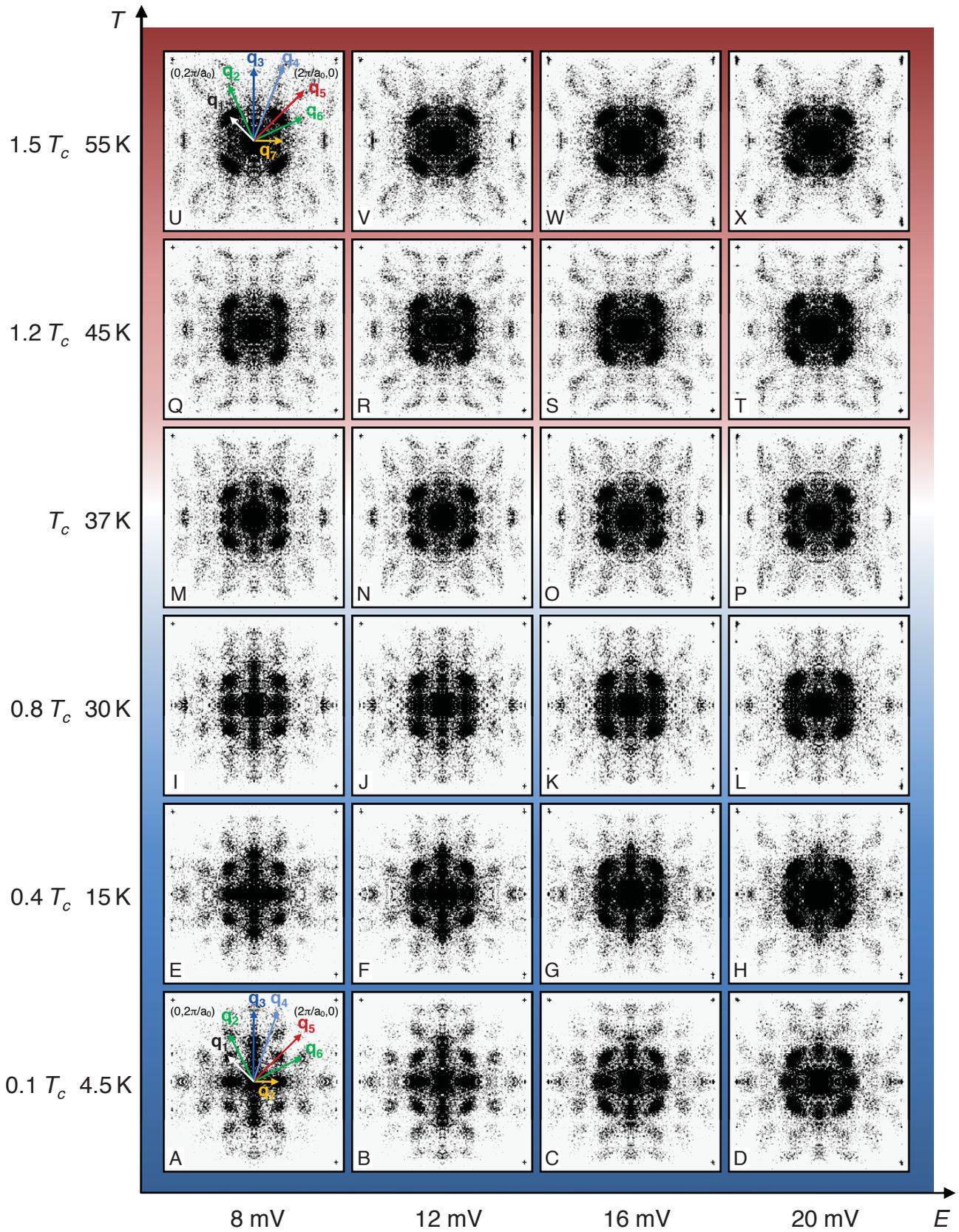

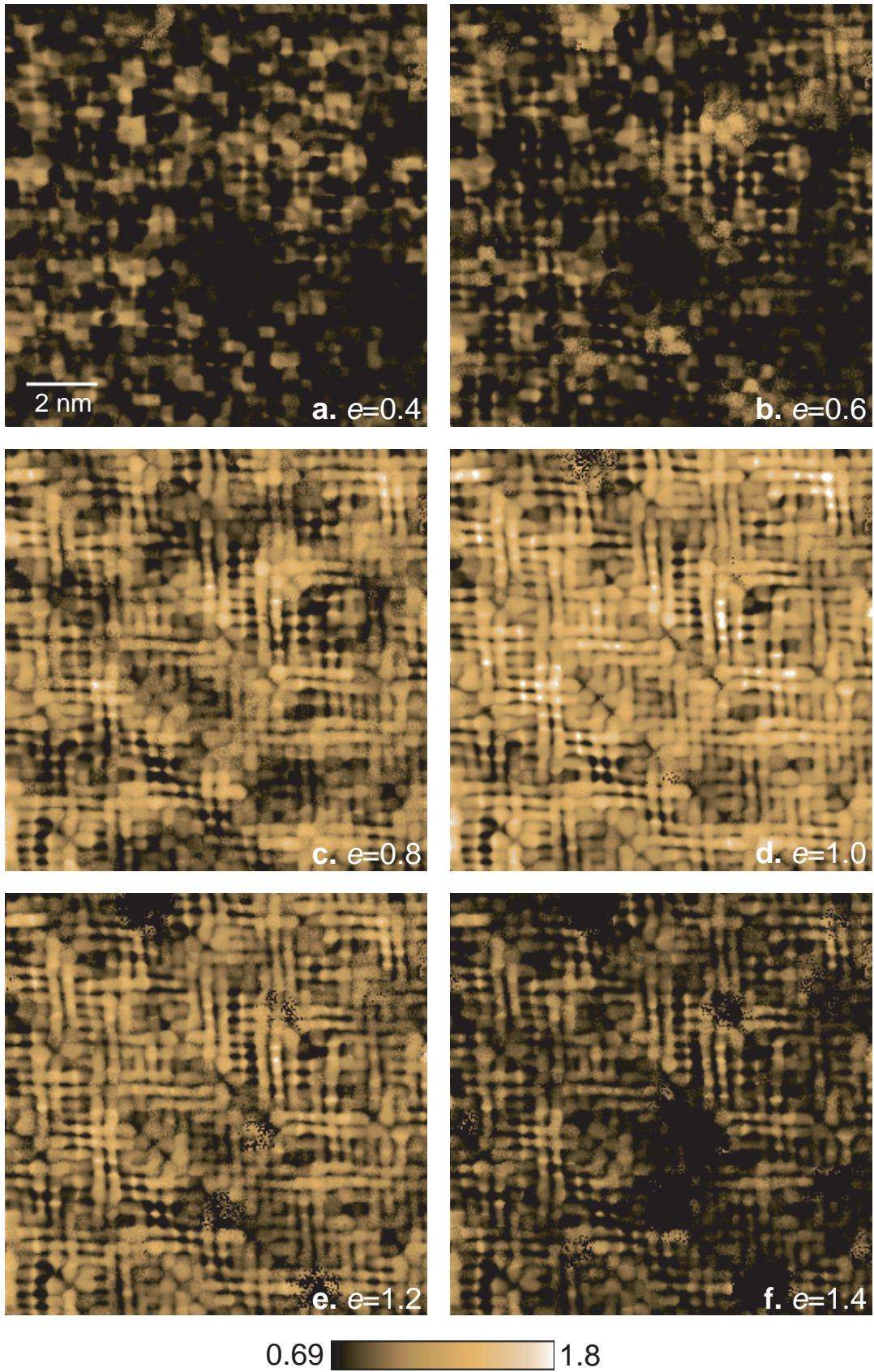

Figure 7

# Figure 8

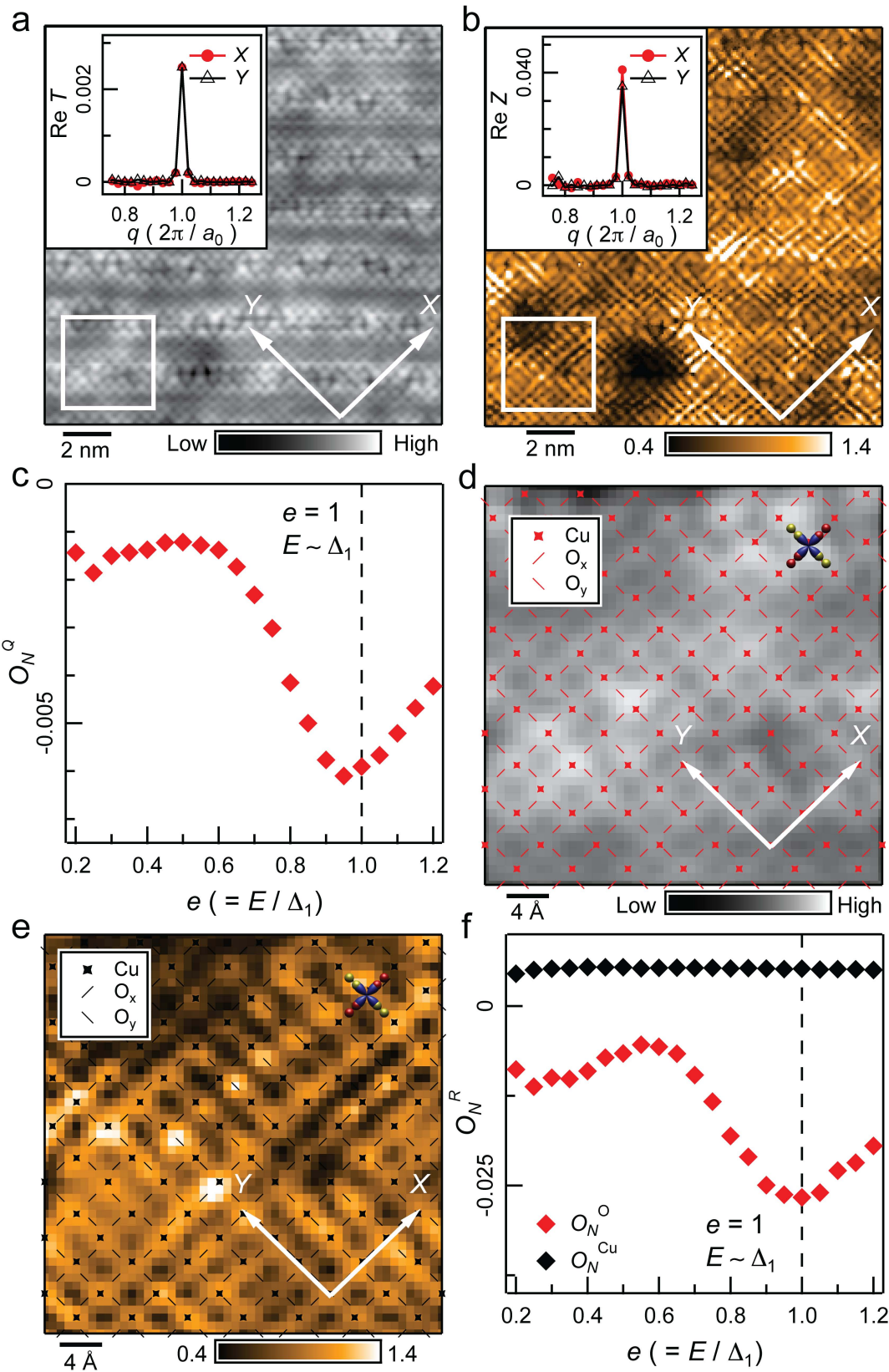

Figure 9

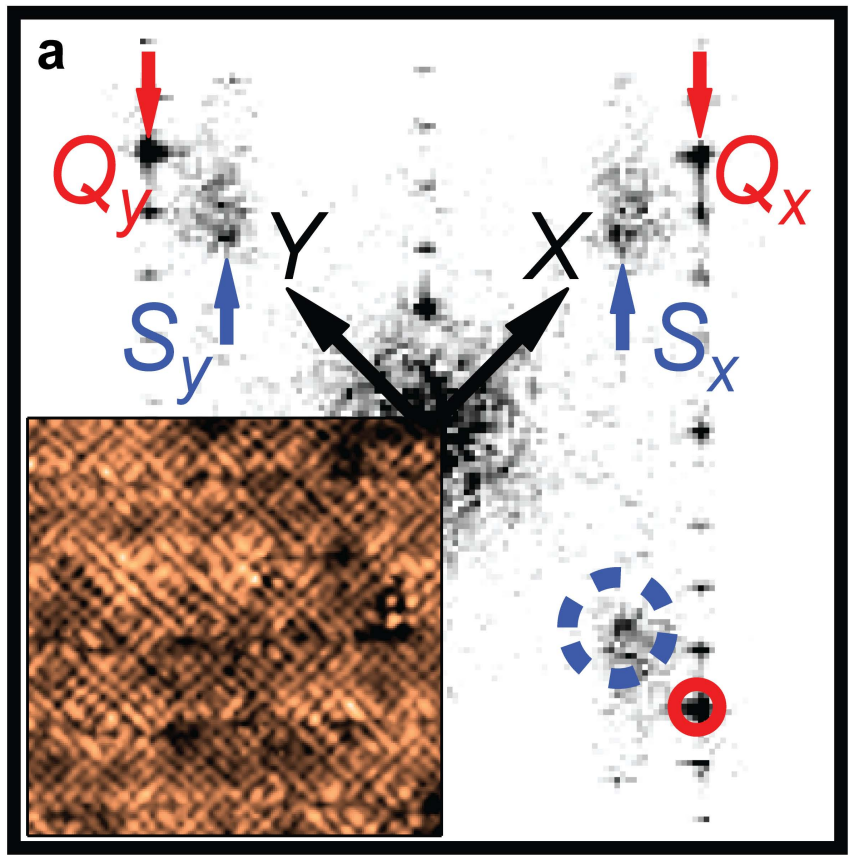
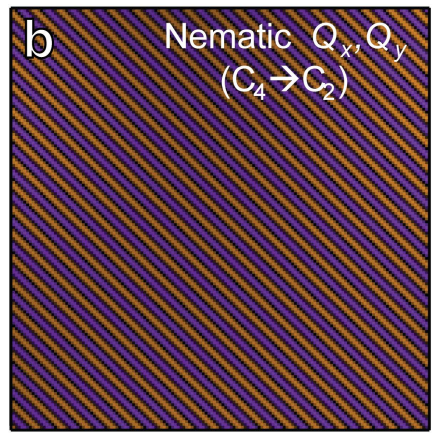
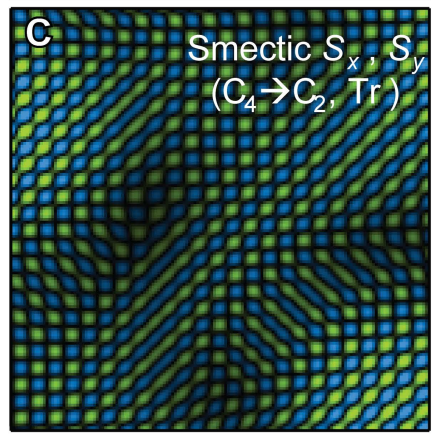

Figure 10

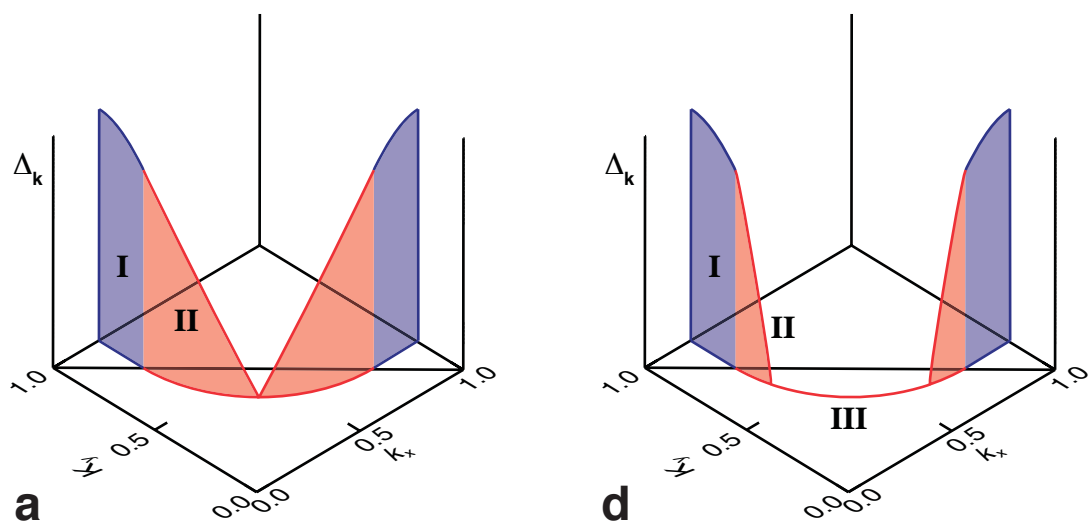

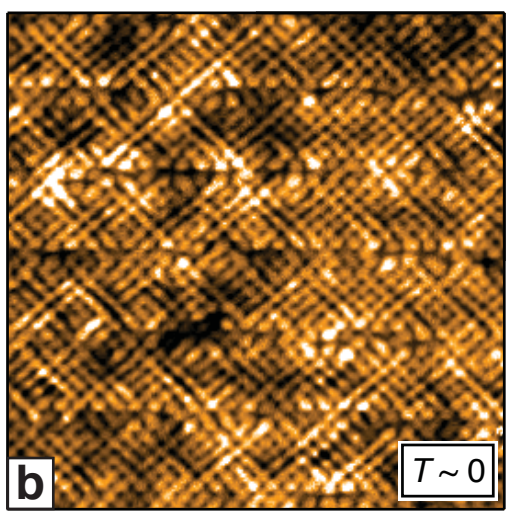
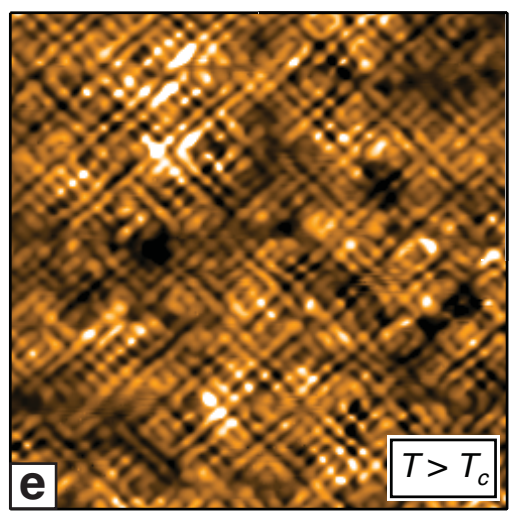

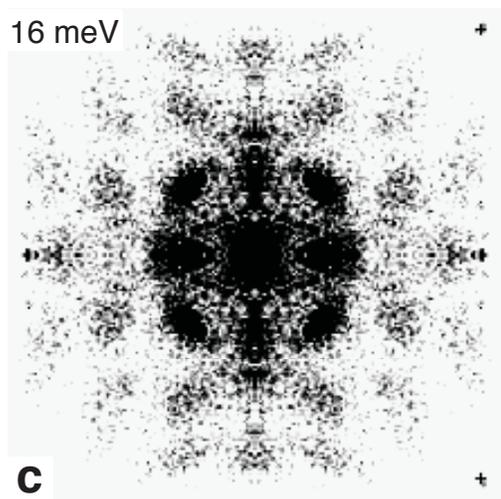
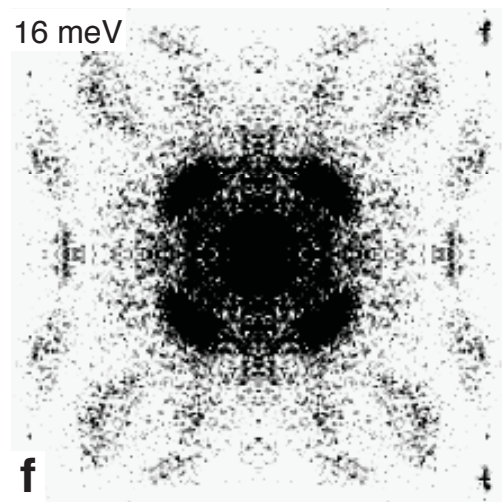

Phase Coherent $d$-SC

Phase Incoherent $d$-SC